\documentclass[a4paper,11pt]{article}
\pdfoutput=1 

\usepackage{jcappub} 
\usepackage[T2A]{fontenc}
\usepackage{booktabs}
\usepackage[english]{babel}
\usepackage{hyperref}

\title{\boldmath Microlensing of dark matter models in the Milky Way}

\author[a,1]{Bichu Li \note{Corresponding author.}}
\author[a]{Chan-Yu Tang}
\author[a]{Zhuo-Ran Huang}
\author[a,2]{Lei-Hua Liu\note{Corresponding author.}}

\affiliation[a]{Department of Physics, College of Physics, Mechanical and Electrical Engineering, Jishou University, Jishou 416000, China}

\emailAdd{libichu@mail.ustc.edu.cn}
\emailAdd{huangzhuoran@126.com}
\emailAdd{liuleihua8899@hotmail.com}

\abstract{We investigate constraints on the abundance of primordial black holes (PBHs) as dark matter (DM) candidates using five years of microlensing data from the OGLE survey. While the majority of OGLE’s $\sim 2000$ microlensing events are well-explained by stellar populations such as brown dwarfs, main-sequence stars, and compact remnants, a subset of six ultrashort-timescale events ($t_E \sim 0.1 - 0.3$ days) may signal the presence of PBHs. Building upon prior work that adopted the Navarro-Frenk-White (NFW) DM profile, we examine how alternative DM halo models---specifically the Einasto and Burkert profiles---affect microlensing predictions and the constraints on PBH abundance. In light of kinematic data of Milky Way, we could obtain the range of ($r_s, \rho_s$) for both profiles. We computed differential microlensing event rates for both profiles, using the main-sequence star rate as an observational benchmark. Our results show that neither the Einasto nor Burkert profiles reproduce the distribution of main-sequence star events, yet both allow for viable explanations of the ultrashort-timescale events with PBH masses $M_{\mathrm{PBH}} \sim 10^{-5} M_\odot$. Using a Poisson likelihood analysis under the null hypothesis that no PBH is observed in OGLE dataset, we derive 95\% C.L. upper and lower bounds on $f_{\mathrm{PBH}}$ for both profiles, finding that the constraints are significantly relaxed under Burkert profiles compared to the NFW case. These results show the sensitivity of PBH constraints to the assumed DM halo structure and highlight the importance of accurately modeling the inner Galactic density profile to robustly assess PBH dark matter scenarios.}

\begin{document}
\maketitle
\flushbottom

\section{Introduction}
\label{sec:intro}
The dark matter (DM) has become an essential component in the standard model of galaxy formation with inflation \cite{Springel:2005nw}. From the perspective of particle physics, Weakly Interacting Massive Particles (WIMPs) represent a viable candidate \cite{Arina:2018zcq, Hooper:2018kfv, Jungman:1995df}, despite the absence of direct detection. Its existence is strongly supported by galactic rotation curves \cite{1980ApJ238471R, Allen:2011zs, Corbelli:1999af, Refregier:2003ct}, galaxy clusters \cite{Allen:2011zs}, and gravitational lensing observations \cite{Refregier:2003ct}.

Another promising candidate is primordial black holes (PBHs), formed through the collapse of overdense regions in the early universe \cite{Carr:1974nx, Hawking:1971ei, Zeldovich:1967lct}. PBHs can form via large overdensities in inflationary models \cite{Cotner:2016cvr, Cotner:2018vug, Inomata:2016rbd}, during preheating \cite{Liu:2021rgq}, etc., spanning a broad mass range including supermassive black holes \cite{Yuan:2023bvh}. Recent PBH research is summarized in Refs.~\cite{Bird:2016dcv, Carr:2025kdk, Sasaki:2016jop, Sasaki:2018dmp}, where PBHs are considered potential binary black hole sources for gravitational wave events detected by LIGO/Virgo \cite{LIGOScientific:2016aoc, LIGOScientific:2017ycc}. However, the PBH fraction in DM is constrained through multiple approaches: black hole evaporation \cite{Boudaud:2018hqb, DeRocco:2019fjq, Laha:2019ssq}, gravitational microlensing \cite{Blaineau:2022nhy, Cai:2022kbp, Delos:2023fpm, EROS-2:2006ryy, Griest:2013aaa, Niikura:2017zjd, Niikura:2019kqi, Oguri:2022fir}, gravitational waves \cite{Hutsi:2020sol, LIGOScientific:2019kan}, dynamical effects \cite{Koushiappas:2017chw}, and cosmic microwave background observations \cite{Poulin:2017bwe, Serpico:2020ehh}.

Microlensing is a powerful tool for exploring the universe \cite{Awad:2023tvu, Kelly:2023mgv}, including investigations of wormholes \cite{Gao:2022cds, Gao:2023ipv, Gao:2023sla, Liu:2022lfb}, etc. It is particularly valuable for studying dark matter in the Milky Way (MW) \cite{MACHO:1990npi, Paczynski:1985jf}. Microlensing causes time-varying magnification of background stars when a lens passes closely through the line-of-sight toward the star. Several microlensing experiments have been conducted, including MACHO \cite{MACHO:2000qbb} and EROS \cite{EROS-2:2006ryy}, which monitored large numbers of stars in the Large Magellanic Cloud (LMC) with 24-hour cadence. The MACHO collaboration ruled out brown dwarfs in the mass range $[10^{-7}, 10]\,M_\odot$ as dark matter candidates \cite{MACHO:2000qbb}, known as massive compact halo objects (MACHOs). The Kepler mission constrained the abundance of $10^{-8}\,M_\odot$ objects through two years of microlensing event data \cite{Griest:2013aaa}. The Optical Gravitational Lensing Experiment (OGLE) collaboration \cite{2015AcA651U, Mrz2017NoLP} has conducted long-term monitoring of millions of stars in Galactic bulge fields for over a decade. These observations have discovered more than 2000 microlensing events and placed significant constraints on exoplanetary systems, brown dwarfs and low-mass stars and free-floating planets in interstellar space \cite{2017Natur548183M}. Similar constraints from the MOA microlensing experiment are presented in Refs.~\cite{Sumi:2002wg, Sumi:2011kj}.

In Ref.~\cite{Niikura:2019kqi}, the authors constrained the abundance of PBHs with microlensing events observed from OGLE, under the assumption of a NFW profile for the DM density distribution and a null hypothesis that no PBH lensing signals are detected in the data. Following their method, we further investigate how dark matter (DM) profiles affect the constraints on \(f_\mathrm{PBH}\). 
In this work, we consider two profiles: the Einasto profile and the Burkert profile. The Einasto profile, proposed in Ref.~\cite{Navarro:2003ew}, provides a better fit to the DM density profile obtained from high-resolution N-body simulations. The Burkert profile \cite{Einasto:1965czb, Graham:2005xx} was introduced to address the core-cusp problem, where the observed rotation curves of dwarf galaxies favor a flat, constant-density core~\cite{Flores:1994gz, Moore:1994yx, deBlok:2001hbg, Shen:2023kkm}, which can not explained by the NFW profile. 


The majority of OGLE microlensing events can be well modeled by astrophysical sources (brown dwarfs, main-sequence stars, and stellar remnants including white dwarfs, neutron stars, and astrophysical black holes).
However, OGLE also reports six ultrashort-timescale events with Einstein timescales $t_{\rm E} \in [0.1, 0.3]~\mathrm{days}$ that deviate from the majority of OGLE observations and are conventionally attributed to free-floating planets—objects, whose formation mechanisms remain poorly understood. Following the suggestion in Ref.~\cite{Niikura:2019kqi}, we will consider the possibility that these ultrashort-timescale events originate from PBH microlensing, and we will use them to constrain the PBH abundance. Under the same null hypothesis as in Ref.~\cite{Niikura:2019kqi}, we find that if the DM halo follows either the Einasto or Burkert profile, the allowed fraction of PBHs in DM is significantly larger than that inferred using the NFW profile. In particular, the tightest upper bound on \( f_{\mathrm{PBH}} \) can be relaxed from the percent level to values over 50\%. This difference can be attributed the varying total enclosed mass within a given radius predicted by different DM profiles.

This paper is structured as follows. Sec. \ref{set up} will establish the Milky Way framework, including our coordinate system and primary targets. Sec. \ref{profile of energy density} will present the density profiles for dark matter lenses, disk lenses, and bulge lenses. Sec. \ref{microlensing basis} will provide foundational microlensing theory. Sec. \ref{optical depth and event rate} will compute differential event rates across all profiles, with particular emphasis on dark matter configurations. Sec. \ref{results} compares predictions from all profiles against five-year OGLE data to constrain PBH masses. Sec. \ref{constraints of PBH abunance} will employ Poisson statistics to derive precise PBH abundance limits. Finally, Sec. \ref{Summary} will present conclusions and future research directions.

\section{Some set up of Milky Way}
\label{set up}

\subsection{The coordinate system}
\label{The coordinate system}
In our calculation, we adopt a Cartesian coordinate system centered at the Galactic center of the Milky Way. The \(x\)-axis points from the Galactic center toward the Earth’s position, the \(y\)-axis lies in the Galactic plane and aligns with the rotation direction of Earth, and the \(z\)-axis is perpendicular to the Galactic disk, following the right-hand rule. In this coordinate system, the position of the Earth is taken to be \((x, y, z) = (8~\text{kpc}, 0, 0)\)~\cite{2016ARA&A..54..529B}.

We focus on microlensing events observed by the OGLE survey~\cite{Mrz2017NoLP}, specifically in the direction specified by Galactic coordinates \((l, b) = (1^\circ.0879, -2^\circ.38)\). For a microlensing event, we denote the distance from the source to the observer (Earth) as \(d\), and the distance from the source to the Galactic center as \(r\). Following the analysis in Ref.~\cite{Niikura:2019kqi}, we assume \(d \approx r\) as both $l$ and $b$ are small, which allows us to express the spatial coordinates of the source as:
\begin{equation}
x = d \cos b \cos l, \quad
y = d \cos b \sin l, \quad
z = d \sin b ~.
\end{equation}
In the conventional spherical coordinate system, these relations correspond to:
\begin{equation}
x = r \sin\theta \cos\phi, \quad
y = r \sin\theta \sin\phi, \quad
z = r \cos\theta ~.
\end{equation}
By comparing the two coordinate systems, we obtain the relation \(\phi = l\) and \(\theta + b = \frac{\pi}{2}\), which implies \(\theta = 92^\circ.38\) and \(\phi = 1^\circ.0879\).

\subsection{Main picture}
\label{general picture}
In this paper, we will investigate the microlensing signatures of dark matter (DM) in the Galactic Center, where the DM density is expected to be significantly higher than in the Galactic disk. The spatial distribution of DM can be described by various density profiles, and it is plausible that a fraction of DM consists of primordial black holes (PBHs). Using microlensing data from the OGLE survey~\cite{Mrz2017NoLP}, we will compute constraints on the PBH mass fraction, $f_{\mathrm{PBH}}$, relative to the total DM content.

In addition to PBHs, microlensing signals can also arise from conventional astrophysical lenses, including main sequence(MS) stars, white dwarfs, brown dwarfs, and neutron stars. The OGLE microlensing dataset captures the combined lensing effects from these populations. Ref.~\cite{Niikura:2019kqi} demonstrates that main sequence stars dominate the microlensing signals in both the Galactic disk and bulge. For comparison, we will also present the microlensing event rate from main sequence stars in our results.

\section{Profile of mass density}
\label{profile of energy density}
The MS stars have different density distributions in the Galactic bulge and disk. For the bulge region, we adopt the bar-shaped stellar density model proposed in Ref.~\cite{1992ApJ387181K}, while for the disk, we use the exponential disk profile described in Ref.~\cite{1986ARA&A24577B}.

For the dark matter (DM) distribution between the Galactic Center and the Earth, there are several widely studied halo profiles, including the Navarro-Frenk-White (NFW) profile~\cite{Navarro:1996gj}, the Einasto profile~\cite{Einasto:1965czb, Graham:2005xx}, and the Burkert profile~\cite{Burkert:1995yz}. The NFW profile has already been considered in the context of PBH microlensing constraints in Ref.~\cite{Niikura:2019kqi}. In this work, we focus on the Einasto and Burkert profiles to compute updated constraints on the PBH abundance.

\subsection{Profile of DM}

\subsubsection{Einasto Profile}
\label{Einasto profile}
\begin{figure}[t]
    \centering
    \includegraphics[width=0.8\linewidth]{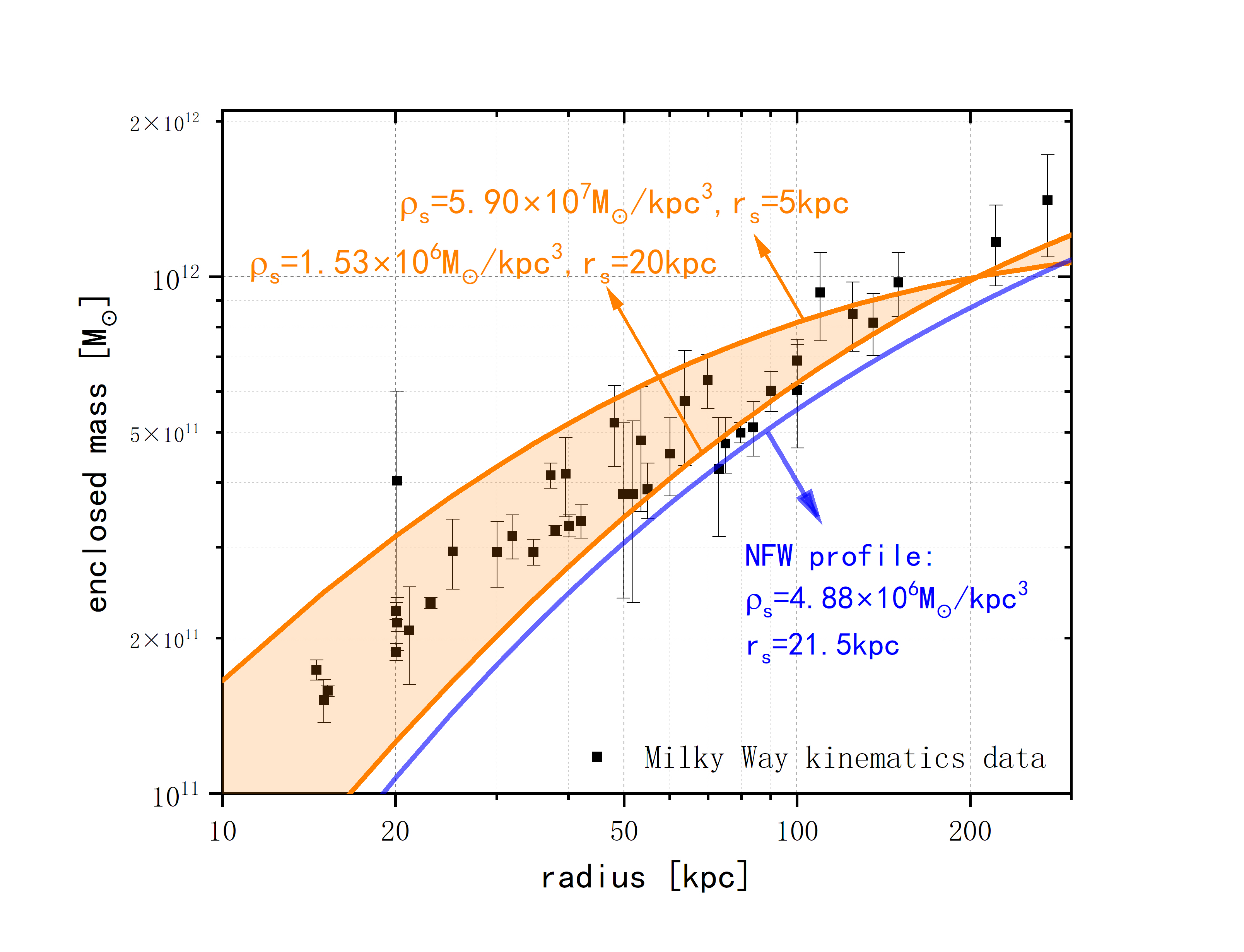}
    \caption{
    The enclosed mass of Einasto profile for Milky Way. The black dots are constraints from kinematic measurements \cite{2025NewAR.10001721H}. Orange curves represent Einasto profile, the blue curve represents NFW profile. The parameters are labeled in the figure. Two Einasto profiles intersect at $r=r_\mathrm{vir}$ as we have assumed the virial mass $10^{12} M_\odot$.
    }
    \label{fig:enclosed_mass_Einasto}
\end{figure}

The Einasto profile was originally introduced in the context of stellar systems \cite{Einasto:1965czb, Graham:2005xx} and has been found to provide an excellent fit to the DM halos in high-resolution numerical simulations~\cite{Navarro:2003ew, Merritt:2005xc}. The density profile is given by
\begin{equation}
    \rho_{\rm DM}^{E}=\rho_s\exp[-\frac{2}{\alpha_s}((r/r_s)^{\alpha_s}-1)],
    \label{profile of Einasto}
\end{equation}
where $\rho_s$ is the scale density and $r_s$ is the scale radius, $\alpha_s=0.17$ characterizes steep the DM profile is. 
Following the approach of Ref.~\cite{Lin:2022rwy}, we fix the total mass of the Milky Way dark matter halo at \( M_\mathrm{vir} = 10^{12} M_{\odot} \) \cite{Niikura:2017zjd}, which corresponds to a virial radius of \( r_\mathrm{vir} = 206 \, \mathrm{kpc} \). This leaves one free parameter between the scale density \(\rho_s\) and scale radius \(r_s\) due to $M_{\rm vir}=\int 4\pi r^2 \rho(r)dr$. To remain consistent with kinematic constraints, we vary \(r_s\) within the range \(5 \, \mathrm{kpc} \le r_s \le 20 \, \mathrm{kpc}\). The resulting enclosed mass profiles are plotted in Fig.~\ref{fig:enclosed_mass_Einasto}, where the shaded region represents the range of Einasto profiles compatible with kinematic data. For comparison, the NFW profile adopted in Ref.~\cite{Niikura:2017zjd} is also shown. Let us explain more about this profile. As $r\ll r_s$, the Einasto Profile \eqref{profile of Einasto} will approach as $\rho(r)\propto \exp[-r^\alpha]$, where it shows the shallow cusp or core-like behavior. Around this scale for the core, the simulation of Einasto Profile will better fit the data compared with NFW profile as $\alpha\approx 0.16-0.2$ \cite{Majumdar:2003mw}. From the FIRE (Feedback In Realistic Environ- ments) project, Ref. \cite{Chan:2015tna} reported that the DM profile is slightly shallower than a NFW profile's prediction within $1~\rm kpc$. Cosmological hydrodynamical simulations of MW sized halo including tuned star formation rate and supernovae feedback evaluated in Ref. \cite{Mollitor:2014ara} will produce the core within $5~\rm kpc$. 

Due to the flexibility of $\alpha$, a superior fit to DM halos in $\Lambda$CDM simulations across all masses (dwarfs to clusters). For $r\gg r_s$, the density \eqref{profile of Einasto} drops faster than NFW profile, its steep outer cutoff may reduce predicted microlensing rates at large galactocentric distances compared to NFW. Finally, the profile is a logarithmic power law, making it analytically tractable for mass and velocity calculations.

\subsubsection{Burkert profile}
\label{Burkert profile}
\begin{figure}[t]
    \centering
    \includegraphics[width=0.8\linewidth]{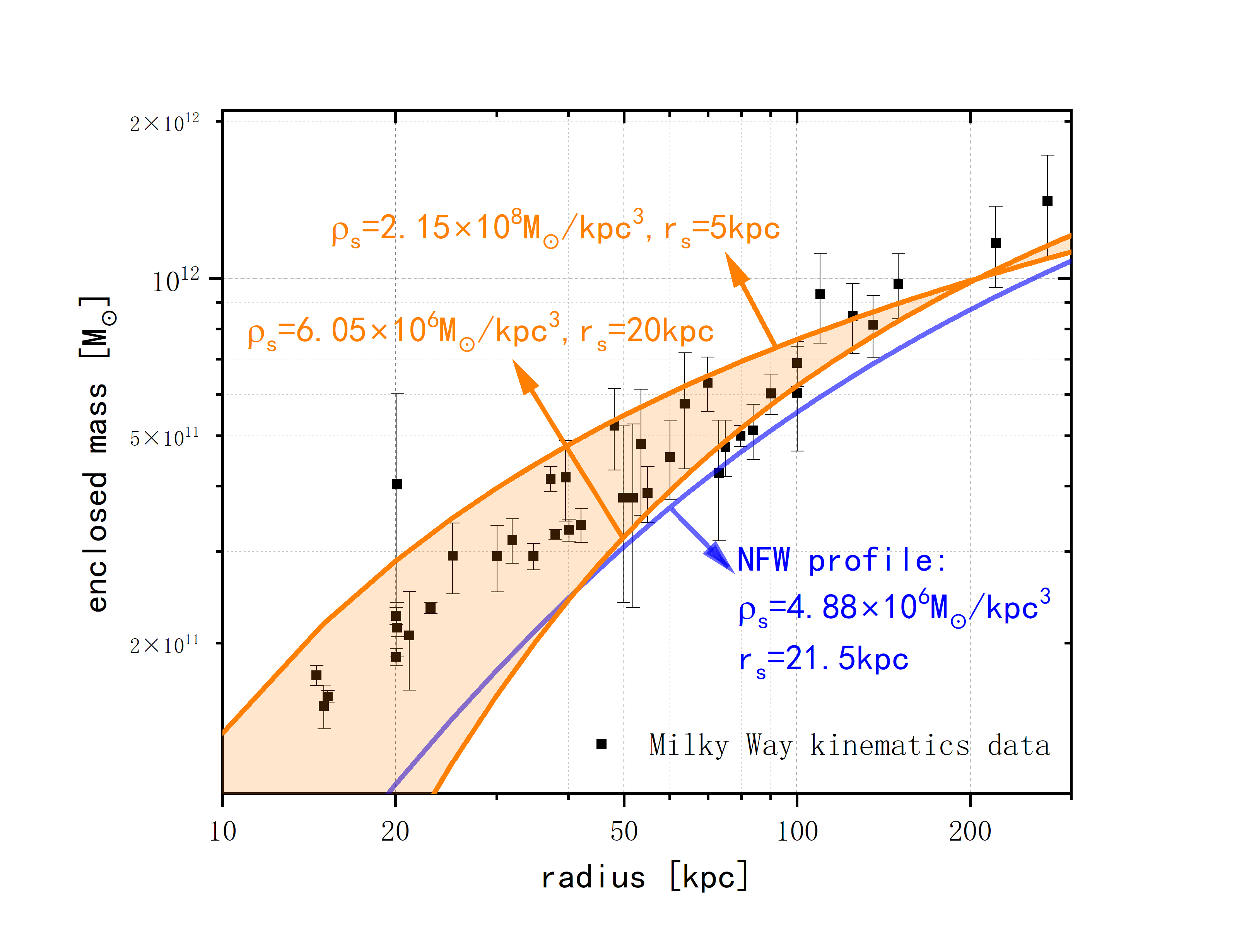}
    \caption{
    The enclosed mass for Burkert profile of Milky Way. The black dots are constraints from kinematic measurements \cite{2025NewAR.10001721H}. Orange curves represent Burkert profile, the blue curve represents NFW profile. The parameters are labeled in the figure. Two Burkert profiles intersect at $r=r_\mathrm{vir}$ as we have assumed the virial mass $10^{12} M_\odot$.
    }
    \label{fig:enclosed_mass_Burkert}
\end{figure}

The Burkert profile is an empirical, cored density profile that was introduced to match the rotation curves of dwarf and low-surface-brightness galaxies~\cite{Burkert:1995yz} and has been shown to provide good fits to observed galaxy rotation curves~\cite{Salucci:2000ps}. The density profile can be written as
\begin{equation}
\rho_{\mathrm{DM}}^B(r) = \frac{\rho_s r_s^3}{(r + r_s)(r^2 + r_s^2)} ~,
\label{profile of Burkert}
\end{equation}
where \(\rho_s\) is the scale density and \(r_s\) is a characteristic scale radius. This profile exhibits a constant density \(\rho \approx \rho_s\) in the central region (\(r \ll r_s\)), transitioning to a steep falloff \(\rho \propto r^{-3}\) at large radii (\(r \gg r_s\)), thus providing a better match to observed galactic dynamics in low-mass systems.
Similar to the last subsection, we also assume a virial mass $M_\mathrm{vir} = 10^{12} M_\odot$ and consider $r_s = 5 \mathrm{kpc}$ and $r_s = 20 \mathrm{kpc}$. The enclosed mass profile are plotted in Fig.~\ref{fig:enclosed_mass_Burkert}.

\subsection{The profile of disk regions}
\label{energy density of disk regions} 
For the mass density distribution of disk region, we adopt the model from
Ref. \cite{1986ARA&A24577B} whose formula is 
\begin{equation}
    \rho_d(R,z)=0.06\times \exp\bigg[-\big(\frac{R-8000}{3500}+\frac{z}{325}\big)\bigg]~M_\odot~\rm pc^{-3},
    \label{mass distribution of disk region}
\end{equation}
where we define \( R^2 = x^2 + y^2 = r \sin \theta \), \( z \) is the direction perpendicular to the Galactic disk (NW disk). As discussed in Section~\ref{The coordinate system}, the approximation \( R \approx r \) remains valid. The disk's mass distribution follows an exponential profile, with a vertical scale height of \( 325 \, \text{pc} \) and a radial scale length of \( 3500 \, \text{pc} \). While the mass-to-light ratio is uncertain, Ref.~\cite{Niikura:2019kqi} normalizes it to \( 0.06 \, M_\odot \, \text{pc}^{-3} \) in the solar neighborhood.

\subsection{The profile of bulge regions}
\label{energy density of bulge regions}
For the stellar population in the bulge regions, we adopt the bar-structured model from \cite{1992ApJ387181K}:
\begin{equation}
    \rho_b(x,y,z)= \begin{cases} 
    1.04 \times 10^6 \left(\frac{s}{0.482\,\text{pc}}\right)^{-1.85} M_\odot \text{pc}^{-3}, & s < 938\,\text{pc}  \\
    3.53 K_{0}\left(\frac{s}{667\,\text{pc}}\right) M_\odot \text{pc}^{-3}, & s \geq 938\,\text{pc}
    \end{cases}   
    \label{eq:density_bulge}
\end{equation}
where $(x, y, z)$ are defined in Section~\ref{The coordinate system}, and $s^4 = R^4 + (z/0.61)^4$ with $R^2 = x^2 + y^2$ representing the distance from the NW region center. Here, $K_0(x)$ is the modified Bessel function, and all coordinates are in parsecs.

The profile shows significant mass concentration within $|s| < 938\,\text{pc}$, becoming sparse beyond this range where most brown dwarfs, main-sequence stars, and stellar remnants reside. Using the coordinate transformations from Section~\ref{The coordinate system}, we express $R = r \sin \theta$, and can approximate $R \approx r$ since $\sin(92^\circ.38) = 0.998773$. Our analysis in Section~\ref{The coordinate system} further shows the $z$-direction contribution is negligible based on velocity distributions.

\section{Microlensing basis}
\label{microlensing basis}
\begin{figure}[t]
    \centering
    \includegraphics[width=0.6\linewidth]{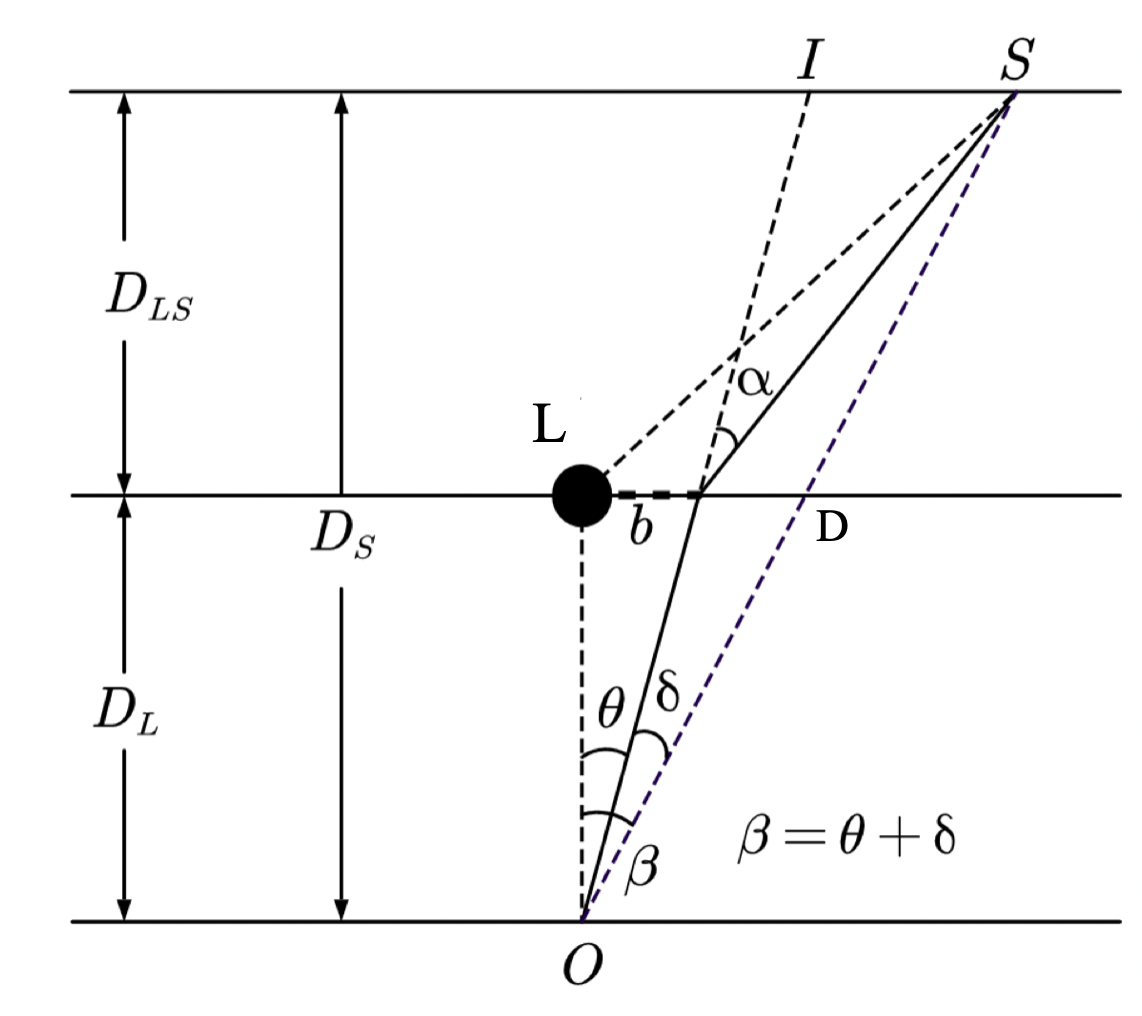}
    \caption{Schematic diagram of gravitational microlensing geometry. $S$ denotes the source, $L$ the lens, and $O$ the observer. $b$ is the impact parameter and $\alpha$ is the deflection angle. Distances between the source, lens, and observer are labeled as $D_S$, $D_L$, and $D_{LS}$, respectively.}
    \label{fig:lensing}
\end{figure}

Microlensing occurs when light from a background source $S$ is deflected by the gravitational field of a foreground object $L$, the lens, as illustrated schematically in Fig.~\ref{fig:lensing}. According to general relativity, the deflection angle $\alpha$ for a point-mass lens is given by~\cite{doi:10.1126/science.84.2188.506, Narayan:1996ba}:
\begin{equation}
\alpha = \frac{4 G M}{c^2 b}~,
\end{equation}
where $M$ is the mass of the lens and $b$ is the impact parameter—the closest distance of the light ray to the lens.

In this work, we focus on microlensing toward the Northwest (NW) region, where potential lensing sources include a DM halo, brown dwarfs, and stellar remnants. As discussed in the previous section, we will consider various density profiles for the DM halo. For continuously distributed lensing mass, the deflection angle generalizes to~\cite{Narayan:1996ba}:
\begin{equation}
\alpha = \frac{4 G M(b)}{c^2 b}~,
\label{general deflection angle}
\end{equation}
where $M(b)$ denotes the mass enclosed within the cylinder of radius $b$. This formalism has also been employed in Refs.~\cite{Fujikura:2021omw, Cai:2022kbp}. The total lensing mass $M_t(b)$ is given by the sum of contributions from different astrophysical populations:
\begin{equation}
M_t(b) = M_{\rm DM}(b) + M_s + M_{sr} + M_b~,
\end{equation}
where $M_{\rm DM}(b)$ is the enclosed DM halo mass, $M_s$ is the mass of main-sequence stars, $M_{sr}$ refers to stellar remnants, and $M_b$ represents brown dwarfs.

For small deflection angles $\alpha, \theta, \delta \ll 1$, the lens equation takes the form~\cite{Schneider:1992bmb, Narayan:1996ba}:
\begin{equation}
\beta = \theta - \frac{D_{LS}}{D_S} \alpha~,
\label{lensing equation}
\end{equation}
where $\beta$ is the angle between lens and source, $\theta$ is the angle between lens and observed image position, $D_L$, $D_S$, and $D_{LS}$ are the distances from observer to the lens, to the source, and between the lens and source, respectively.

When the source lies directly behind the lens ($\beta = 0$), the image forms a symmetric ring known as the Einstein ring, characterized by the Einstein angle $\theta_E$:
\begin{equation}
\theta_E = \frac{D_{LS}}{D_S} \frac{4 G M}{c^2 b}~.
\end{equation}
Using $b = D_L \theta$, one obtains the physical Einstein radius on the lens plane:
\begin{equation}
r_E = \theta_E D_L = \frac{2}{c} \sqrt{\frac{D_{LS} D_L}{D_S} G M}~.
\end{equation}

The Einstein radius serves as a scale that determines whether a lens can be approximated as a point. In particular, the DM halo’s scale radius $r_* = 21.5~\mathrm{kpc}$ is significantly larger than its Einstein radius ($r_* > r_E$), therefore it's necessary to employ a continuous density profile. On the other hand, for compact objects like stars, brown dwarfs, and remnants, their physical sizes are negligible compared with $r_E$ and the point-mass approximation remains valid~\cite{Niikura:2019kqi}.

The duration of a microlensing event is typically quantified by the Einstein time $t_E$, defined as the time taken by the lens object to traverse a distance equal to $r_E$:
\begin{equation}
t_E = \frac{r_E}{v} = \frac{2}{cv} \sqrt{G M_t \frac{D_{LS} D_L}{D_S}}~,
\label{einstein time}
\end{equation}
where $v$ is the projected relative velocity between the source and lens on the lensing plane. The above equation can also be written as \cite{Niikura:2019kqi}:
\begin{equation}
t_E \simeq 44~\mathrm{days} \left( \frac{M_t}{M_\odot} \right)^{1/2} \left( \frac{D}{4~\mathrm{kpc}} \right)^{1/2} \left( \frac{v}{220~\mathrm{km/s}} \right)^{-1}~,
\label{times scaling}
\end{equation}
where $D = D_L D_{LS} / D_S$ is the effective lensing distance. It can be seen that, for fixed effective lensing distance $D$ and transverse velocity $v$, the Einstein timescale increases with the square root of the lens mass. The OGLE survey reports microlensing event with durations ranging from $0.1$ to $300$ days, which correspond to lens masses spanning approximately $10^{-6}~M_\odot$ to $10~M_\odot$~\cite{Niikura:2019kqi}, covering a diverse population from sub-Earth-mass objects to stellar-mass black holes.

Since $v$ varies for different lensing events, we describe it statistically. The transverse velocity is decomposed as:
\begin{equation}
v^2 = v_y^2 + v_z^2~,
\label{decomposition of velocity}
\end{equation}
with $v_y = v \cos\theta$ and $v_z = v \sin\theta$ are two orthogonal velocity components on the lensing plane. Assuming $v_y$ and $v_z$ are independent and normally distributed, the velocity distribution $f(v)$ becomes:
\begin{equation}
f(v) = f(v_y) f(v_z)~,
\end{equation}
where
\begin{equation}
f(v_y) = \frac{1}{\sqrt{2\pi} \sigma_y} \exp\left( -\frac{(v_y - \bar{v}_y)^2}{2 \sigma_y^2} \right), \qquad
f(v_z) = \frac{1}{\sqrt{2\pi} \sigma_z} \exp\left( -\frac{v_z^2}{2 \sigma_z^2} \right)~.
\end{equation}
Following Ref.~\cite{Niikura:2019kqi}, we adopt $\bar{v}_z = 0$. The parameters $\bar{v}_y$, $\sigma_y$, and $\sigma_z$ depend on the specific population (disk, halo, bulge) and will be specified in our numerical setup. To elaborate on the velocity distribution formalism, we assume that stellar components are kinematically supported exclusively by isotropic velocity dispersion. Consequently, the velocity distribution contains no rotational component. 

\section{Optical depth and event rate}
\label{optical depth and event rate}
In this section, we introduce the Einstein radius \( r_E \) and timescale \( t_E \) to define the microlensing optical depth and event rate. The optical depth \(\tau\) represents the instantaneous probability that a source star lies within the Einstein radius of a foreground lensing object, corresponding to a magnification threshold of \( A > 1.43 \). 

The total optical depth is the sum of contributions from different Galactic components:
\begin{equation}
    \tau = \tau_b + \tau_d + \tau_{\rm DM},
    \label{eq:optical_depth}
\end{equation}
where \(\tau_b\) and \(\tau_d\) denote the optical depths for the bulge and disk regions, respectively, while \(\tau_{\rm DM}\) accounts for dark matter (primarily primordial black holes, PBHs) in the Galactic center. 

To connect with observational data (e.g., OGLE), we define the differential event rate as the frequency of microlensing events per source star per unit observational time \( t_{\rm obs} \) for a given timescale \( t_E \):
\begin{equation}
    \frac{d\Gamma_a}{dt_E} = \frac{d^2\tau_a}{dt_{\rm obs} \, dt_E},
    \label{eq:diff_event_rate}
\end{equation}
where \(\Gamma_a\) is the event rate, and the subscript \( a \) labels the lens population (\( b \): bulge, \( d \): disk, \( \rm DM \): dark matter). Equations \eqref{eq:optical_depth} and \eqref{eq:diff_event_rate} form the foundational framework for our analysis. All of the optical depth and differential event rate comes from \cite{Niikura:2019kqi}. 

\subsection{PBH lensing}
\label{pbh lensing}
Following the treatment from Ref.~\cite{Niikura:2019kqi}, we adopt a single source plane approximation, i.e., assume \( D_S = 8  \text{kpc} \) as the distance to the source stars in the bulge region. If dark matter (DM) consists partially of primordial black holes (PBHs), microlensing events can occur when a PBH passes through the line of sight from observer to the background star.

We assume a \textbf{monochromatic PBH mass function} and model the DM spatial distribution using both the Einasto profile (Eq. \eqref{profile of Einasto}) and Burkert profile (Eq. \eqref{Burkert profile}). Using the coordinate system defined in Sec. \ref{The coordinate system}, the optical depth for the DM halo is given by:

\begin{equation}
    \tau_{\mathrm{DM}} = \frac{4\pi G}{c^2} \int_{0}^{D_S} \rho_{\mathrm{DM}}(D_L)  \frac{D_L D_{LS}}{D_S}  dD_L,
    \label{optical depth of DM halo}
\end{equation}
where \(\rho_{\mathrm{DM}}\) is the DM mass density profile, and \(D_L\), \(D_S\), and \(D_{LS}\) are the lens, source, and lens-source distances, respectively (defined in Fig. \ref{fig:lensing}). \textbf{When PBHs constitute only a fraction \(f_{\mathrm{PBH}}\) of the DM, the PBH optical depth is modified to}:
\begin{equation}
  \tau_{\mathrm{PBH}} = f_{\mathrm{PBH}} \cdot \tau_{\mathrm{DM}}. 
  \label{optical depth for pbh}
\end{equation}
Regarding the differential event rate in the bulge region, it is expressed as:
\begin{equation}
    \frac{d\Gamma_{\mathrm{DM}}}{dt_E} = \pi 
    \int_0^{\bar{D}_S} d{D_L} 
    \frac{\rho_{\mathrm{DM}}}{M_{\mathrm{PBH}}} 
    \int_{-\pi/2}^{\pi/2} d{\theta} 
    v_{\perp}^4 f_{\mathrm{DM}}(v_{\perp}, \theta),
    \label{differential event rate for pbh}
\end{equation}
where \( f_{\mathrm{DM}}(v_{\perp}, \theta) \) denotes the velocity distribution of PBHs. For DM, we will set he mean relative velocity for a PBH lens as follows,
\begin{equation}
    \bar{v}_{\rm PBHy}=-220(1-\alpha)~km/s, \bar{v}_{\rm PBHz}=0, 
    \label{mean velocity of pbh}
\end{equation}
where $\alpha=D_L/D_{LS}$. As for the the velocity dispersion of PBH, we could also have 
\begin{equation}
    \sigma_{\rm PBHy}^2=\sigma_{D }^2+\alpha^2100^2 \rm (km/s^2),~~\sigma_{\rm PBHz}^2=\sigma^2_{\rm DM}+\alpha^2 100^2 \rm (km/s)^2,
    \label{dispersion relation of pbh}
\end{equation}
where $\sigma_{\rm DM}=220~\rm km/s$. 

\subsection{Disk region lensing}
\label{disk lensing}
We now derive the microlensing event rate for disk-region stellar lenses acting on bulge-region source stars. This calculation parallels the bulge-lens case treated in Section 3, with modifications for the disk geometry. Adopting a single-source-plane approximation for computational efficiency similarly to the preceding subsection, we fix all source stars at the Galactic center distance $\bar{D}_S = {8}{\rm ~kpc}$.

The optical depth for disk lenses is given by:
\begin{equation}
\tau_d = \frac{4\pi G}{c^2} \sum_i \int_0^{\bar{D}_s} 
    \rho_{d,i}(d_L) \frac{D_L D_{LS}}{D_S} d{D_L},
\end{equation}
where:
\begin{itemize}
    \item $\rho_{d,i}(D_L)$ is the mass density profile of the $i$-th stellar component
    \item $D_L$ denotes the lens distance along the line of sight
    \item $\bar{D}_s =8~\rm kpc$ is the mean source distance
    \item $D_{\mathrm{LS}} = {D}_S - D_L$ is the lens-source separation
\end{itemize}

Similarly, the differential event rate is:
\begin{equation}
\frac{d\Gamma_d}{dt_E} = \pi \sum_i \int_0^{\bar{d}_s} d{D_L} \frac{\rho_{d,i}(D_L)}{M_i} \int_{-\pi/2}^{\pi/2} d{\theta}  v_{\perp}^4 f_{d,i}(v_{\perp},\theta),
\label{differential event rate for disk region}
\end{equation}
where \(v_{\perp} = 2R_E \cos \theta / t_E\), and \(f_{d,i}(v_{\perp},\theta)\) denotes the velocity distribution for components perpendicular to the \textbf{line-of-sight} direction for the \(i\)-th stellar component in the disk region.

The mean velocity for the disk region can read as 
\begin{equation}
    \bar{v}_{dy}=220~\rm km/s,~~\bar{v}_{dz}=0,
    \label{mean velocity of disk region}
\end{equation}
For the velocity dispersion, it is assumed followed via \cite{Han:1995zy}, whose resulting formulas are 
\begin{equation}
    \sigma_{dy}^2=(\kappa\delta+30)^2+100\alpha^2~(\rm km/s)^2,~~ \sigma_{dz}^2=(\lambda\delta+30)^2+100\alpha^2~(\rm km/s)^2
\end{equation}
with 
\begin{equation}
    \kappa\equiv5.625\times10^{-3}\rm km/s/pc,~\lambda\equiv 3.75\times 10^{-3}\rm km/s/pc, ~\delta=(8000-x)\rm pc,
    \label{values of parameters}
\end{equation}
where $\kappa$ and $\lambda$ are the dispersion gradient coefficients.  

\subsection{Bulge region lensing}
\label{bulge region lensing}
In this section, we will consider the source and lens are both located in the bulge region. The optical depth for bulge lenses is given by 
\begin{align}
\tau_b &\equiv \frac{1}{N_s} \int_{d_{s,\min}}^{d_{s,\max}} dD_S  n_s(D_S) 
       \sum_i \int_{D_{s,\min}}^{D_s} d{D_L} 
       \frac{\rho_{b,i}(D_L)}{M_i} \pi R_E^2(M_i) \\
&= \frac{4\pi G}{c^2 N_s} \int_{d_{s,\min}}^{d_{s,\max}} d{D_S}  n_s(D_S) 
   \sum_i \int_{D_{S,\min}}^{D_S} d{D_L} 
   \rho_{b,i}(D_i) D,
\end{align}
where:
\begin{itemize}
    \item $D_S$ and $D_L$ are source and lens distances respectively
    \item $D_{s,\min}$, $D_{s,\max}$ define the source distance range
    \item $n_s(D_S)$ is the source star density distribution
    \item $\rho_{b,i}(D_L)$ is the mass density of the $i$-th bulge component
    \item $M_i$ is the mass of lens objects in component $i$
    \item $R_E$ is the Einstein radius
    \item $D$ represents the geometric factor $\frac{D_L D_{LS}}{D_S}$
    \item $N_s$ represents the surface number density of source stars defined by a line-of-sight integration of the three- dimensional number density distribution of source stars, whose formula is $\int_{D_{s,mim}}^{D_{s,max}}dD_S n_s(D_S)$. 
\end{itemize}
The corresponding differential event rate is 
\begin{equation}
\frac{d\Gamma_b}{dt_E} = \pi \sum_i \int_0^{\bar{D}_s} d{D_L} \frac{\rho_{b,i}(D_L)}{M_i} \int_{-\pi/2}^{\pi/2} d{\theta}  v_{\perp}^4 f_{b,i}(v_{\perp},\theta),
\label{differential event rate for bulge region}
\end{equation}
The mean the transverse velocities is
\begin{equation}
    \bar{v}_{by}=-220(1-\alpha)~ \rm km/s, ~\bar{v}_{bz}=0,
    \label{mean velocity of bulge region}
\end{equation}
its corresponding velocity dispersion are given by 
\begin{equation}
    \sigma_{by}^2=(1+\alpha)^2 (100~\rm km/s)^2,~\sigma_{bz}^2=(1+\alpha)^2 (100~\rm km/s)^2. 
    \label{disper relation of bugle region}
\end{equation}

\section{Results}
\label{results}
\subsection{OGLE data}
\label{ogle data}
In this work, all numerical data are based on the OGLE-IV sky survey conducted over 5 years (2011--2015)~\cite{2015AcA651U,2017Natur548183M}. These surveys were performed using the $1.3\,\mathrm{m}$ Warsaw Telescope located at Las Campanas Observatory, Chile. The OGLE project carried out long-term monitoring observations of nine fields in the Galactic bulge region with cadences of either $20\,\mathrm{min}$ or $60\,\mathrm{min}$, covering a total area of $12.6\,\mathrm{square\ degrees}$. 

Through careful observations, they identified and characterized $2622$ microlensing events in terms of the Einstein timescale $t_{\mathrm{E}}$. The substantial volume of data enables stringent constraints on the abundance and mass distribution of various lensing object populations. Refs.~\cite{2017Natur548183M,Niikura:2019kqi} show that the primary contributions to the OGLE data come from main-sequence stars, brown dwarfs (BDs), white dwarfs (WDs), \textit{etc.} All data used in this analysis are taken from~\cite{2017Natur548183M}. In this paper, we will constrain the abundance of PBHs $f_{\mathrm{PBH}}$ using the event rate with various profiles. 

\subsection{Numeric of event rate}
\label{numeric of event rate}
In this section, we investigate the differential event rates of lensing contributions from PBHs with monochromatic mass spectrum, the Galactic bulge, and the Galactic disk, assuming the Burkert profile (Eq.~\eqref{Burkert profile}) and the Einasto profile (Eq.~\eqref{Einasto profile}) distributions for DM, 
and $f_{\rm PBH} = 1$ (i.e., PBHs constitute all dark matter). 
As discussed in Sec.~\ref{general picture}, the lensing contribution from main-sequence stars dominates the observed microlensing events in OGLE data. For the Burkert and Einasto profiles, the differential event rate of main-sequence stars will serve as the observational benchmark. 

\subsubsection{Event rate for Einasto profile}
\label{event rate for Einasto profile}

\begin{figure}[t]
    \centering
    \includegraphics[width=0.9\linewidth]{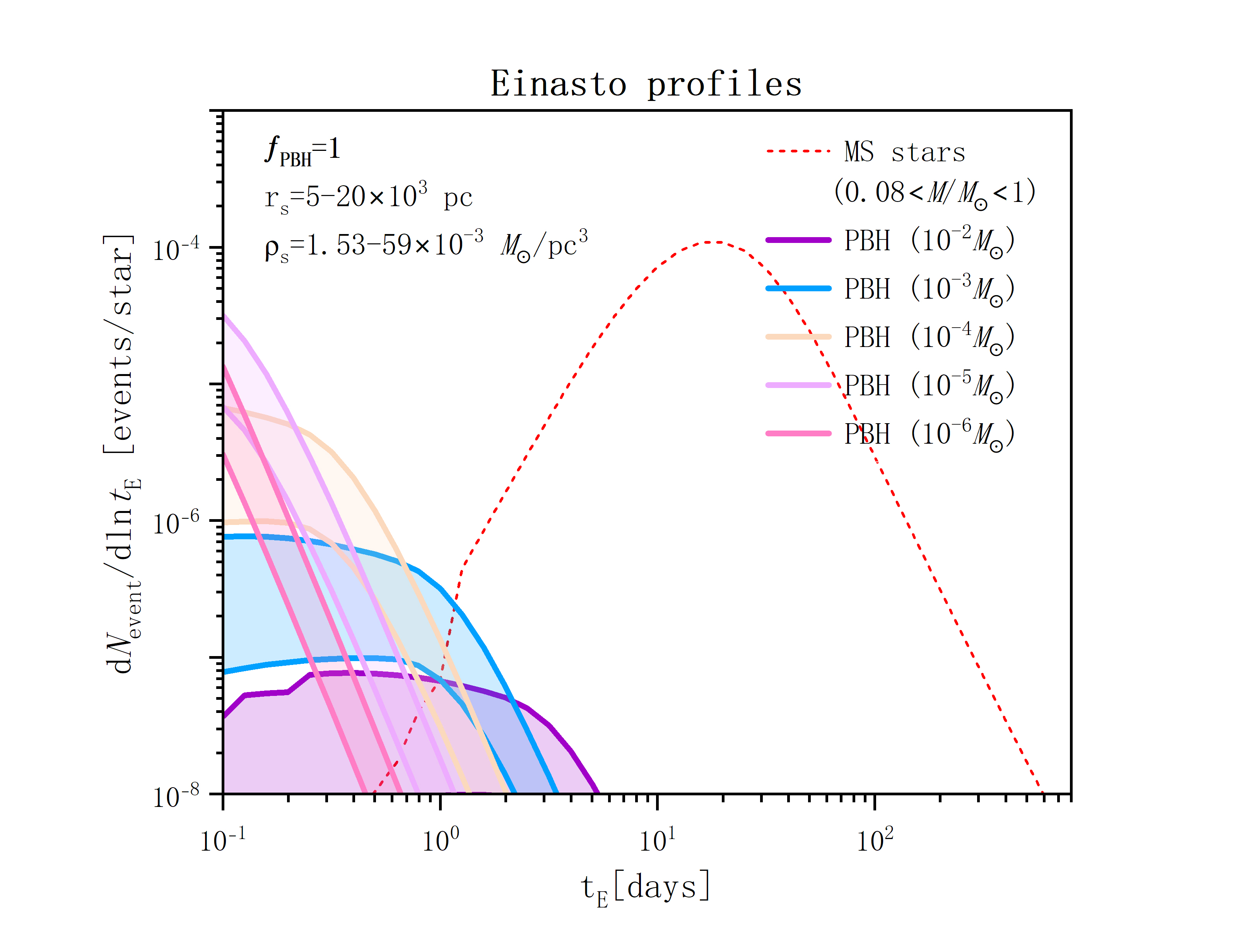}
    \caption{The expected differential number of microlensing events per logarithmic interval of the light curve timescale $t_E$ for the Einasto profile, where there is only one single star in the bulge region. And we have assumed that the five years data of the OGLE data for the MS stars. The quantity of $\frac{dN_{\rm event}}{d\ln t_{\rm E}}=5\times \rm years\times \frac{d\Gamma_a}{dt_{\rm E}}$, including Eq. \eqref{differential event rate for pbh}, Eq. \eqref{differential event rate for disk region} and Eq. \eqref{differential event rate for bulge region}. The other parameters are set in the figure. 
    For each PBH mass, the lower bound corresponds to the profile with a lower scale density, \(\rho_s = 1.53 \times 10^{-3} \, M_\odot \, \mathrm{pc}^{-3}\), and a larger characteristic radius, \(r_s = 20 \times 10^{3} \, \mathrm{pc}\). On the other side, the upper bound represents the profile with a higher scale density, \(\rho_s = 59 \times 10^{-3} \, M_\odot \, \mathrm{pc}^{-3}\), and a smaller characteristic radius, \(r_s = 5 \times 10^{3} \, \mathrm{pc}\).
    }
    \label{fig:differential event rate fo Einasto}
\end{figure}

The microlensing events from the MS stars have a different event rate at disk and bulge region, given by \ref{differential event rate for disk region} and \ref{differential event rate for bulge region} respectively. Ref.~\cite{Niikura:2019kqi} has shown that MS stars are the dominant contributors to the microlensing events observed in the OGLE dataset. Therefore we adopt the differential event rate from MS stars as a reference benchmark for the results of PBHs. We calculated the expected number of events per logarithmic interval of Einstein timescale $t_E$. For a 5-year observation period, this quantity is given by \cite{Niikura:2019kqi}:
\begin{equation}
\frac{dN_{\rm event}}{d\ln t_E} = 5\,\mathrm{yr} \times \frac{d\Gamma_a}{dt_E},
\end{equation}
where $d\Gamma_a/dt_E$ denotes the differential event rate (Eq.~\ref{differential event rate for pbh}).

The results are shown in Figure~\ref{fig:differential event rate fo Einasto}. The red dashed line shows the differential event rate for MS stars, while the shaded regions represent the differential event rates for PBHs of different masses. For each PBH mass, the upper bound corresponds to the parameter choice with highest scale density, \(\rho_s = 59 \times 10^{-3} \, M_\odot \, \mathrm{pc}^{-3}\). Assuming $f_{PBH}=1$, a higher scale density implies a larger number of PBHs of a given mass, resulting in a higher event rate. It can be seen that, assuming the Einasto profile for DM distribution, PBH can not reproduce the differential event rates of MS stars. Either the position or the amplitude of the peak would not match the prediction from MS stars. For $f_\mathrm{PBH}=1$, i.e., the entire dark matter content consists of PBHs, and assume a monochromatic mass spectrum, a lower bound for PBH mass can be found at $10^{-3}M_{\odot}$. The lensing signal of PBHs with a smaller mass will dominant over MS stars, contradicting OGLE observations. Similar mass constraints under the monochromatic assumption have been discussed in the context of other microlensing surveys~\cite{Niikura:2017zjd, Carr:2020xqk}.

\subsubsection{Event rate for Burkert Profile}
\label{event rate for Burkert profile}
\begin{figure}[t]
    \centering
    \includegraphics[width=0.9\linewidth]{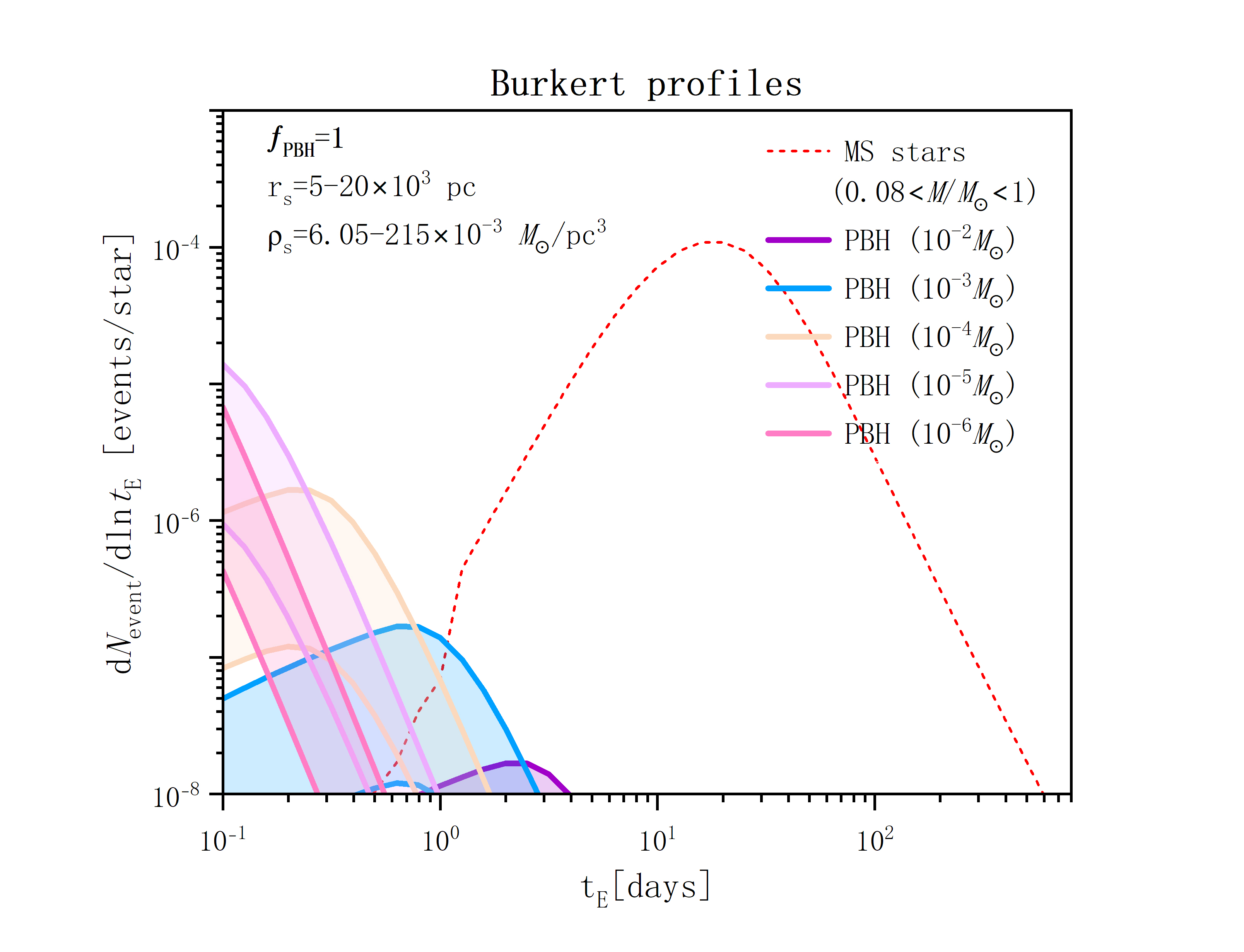}
    \caption{Expected number of microlensing events per logarithmic interval of the Einstein timescale \( t_E \) for the Burkert dark matter profile, assuming a single source star located in the Galactic bulge. Specifically, the plotted quantity is defined as$\frac{dN_{\mathrm{event}}}{d\ln t_E} = 5 \times \mathrm{years} \times \frac{d\Gamma_a}{dt_E}$, corresponding to a 5-year observation period, consistent with the OGLE dataset. Shaded regions represent the model predictions for PBHs with a monochromatic mass distribution, while dashed curves show results for MS stars with masses in the range \( 0.08 < M/M_\odot < 1 \). The differential event rates \( \frac{d\Gamma_a}{dt_E} \) for lens populations in the disk, bulge, and PBH scenarios are given by Eqs.~(\ref{differential event rate for disk region}), (\ref{differential event rate for bulge region}), and (\ref{differential event rate for pbh}), respectively. For each PBH mass, the lower bound corresponds to the profile with a lower scale density, \(\rho_s = 6.05 \times 10^{-3} \, M_\odot \, \mathrm{pc}^{-3}\), and a larger characteristic radius, \(r_s = 20 \times 10^{3} \, \mathrm{pc}\). On the other side, the upper bound represents the profile with a higher scale density, \(\rho_s = 215 \times 10^{-3} \, M_\odot \, \mathrm{pc}^{-3}\), and a smaller characteristic radius, \(r_s = 5 \times 10^{3} \, \mathrm{pc}\).
    }
    \label{fig:event_rate_for_Burkert_profile}
\end{figure}

Following the method of the previous subsection, we compute the microlensing event rate for dark matter with a Burkert profile. 
The results are presented in Fig.~\ref{fig:event_rate_for_Burkert_profile}.
Combining with the results from last subsection, some important features can be summarized as following:

\begin{itemize}
    \item \textbf{Mass-velocity scaling:} Equation~\ref{einstein time} indicates that $t_E \propto M^{1/2} v^{-1}$. For our monochromatic PBH mass spectrum with Gaussian velocity distribution, the event rate peak corresponds to the maximum likelihood of relative velocity, i.e., the mean relative velocity $\bar{v}_{\rm PBH}$ (Eq.~\ref{mean velocity of pbh}).
    
    \item \textbf{Rightward shift:} For a fixed velocity distribution, increasing $M_{\rm PBH}$ will cause the peak to shift rightwards, as shown in Fig.~\ref{fig:event_rate_for_Burkert_profile}.
    
    \item \textbf{Amplitude suppression:} A higher PBH mass means a lower number density ($n_{\rm PBH} \propto \rho_{\rm DM}/M_{\rm PBH}$), and a lower number density means a lower event rate. Therefore increasing PBH mass will suppress the overall amplitude of differential event rate.
\end{itemize}

Suppose that the DM halo follows the Burkert profile, Fig.~\ref{fig:event_rate_for_Burkert_profile} demonstrate that PBHs still cannot constitute all DM. Similar to the case discussed in the previous subsection for the Einasto profile, the differential microlensing event rates predicted for PBHs fail to reproduce the shape and magnitude of the observed event rates attributed to main-sequence stars when \( f_{\mathrm{PBH}} = 1 \).

\subsection{Comparison with OGLE data}
\label{comparison with ogle data}
\begin{figure}[t]
    \centering
    \includegraphics[width=0.6\linewidth]{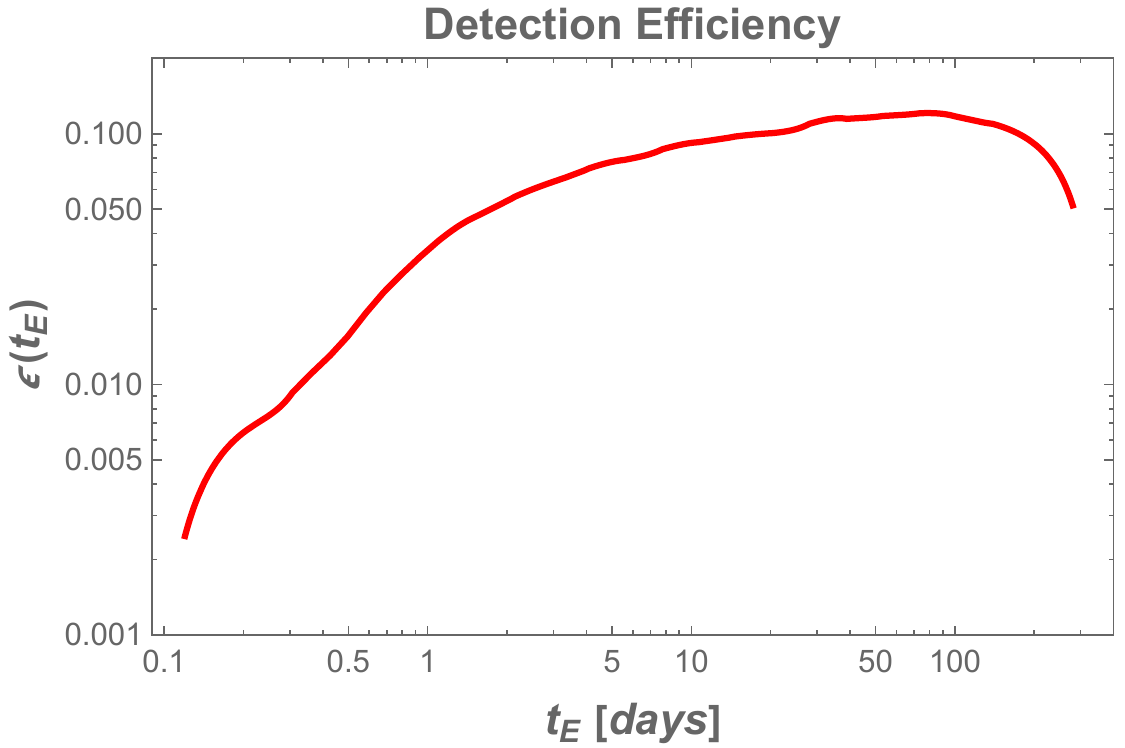}
    \caption{Detection efficiency as a function of event timescale $t_{\rm E}$, derived from \cite{2017Natur548183M}. The curve represents the probability of detecting microlensing events at different timescales in the OGLE survey.}
    \label{fig:detection_efficiency}
\end{figure}

The five-year OGLE data includes $2,\!622$ events for light-curve timescales $t_{\rm E} \in [10^{-1}, 300]$ days. To connect theory with observations, we define the expected number of microlensing events per timescale as:
\begin{equation}
    N_{\rm exp}(t_{\rm E}) = t_{\rm obs} N_s f_A \int_{t_{\rm E}-\Delta t_{\rm E}/2}^{t_{\rm E}+\Delta t_{\rm E}/2} \! d\ln t'_{\rm E} \, \frac{d^2\Gamma}{d\ln t_{\rm E}} \epsilon(t_{\rm E}),
    \label{eq:expected_number}
\end{equation}
where:
\begin{itemize}
    \item $t_{\rm obs}$ is the total observation time
    \item $N_s$ is the number of source stars in the OGLE survey
    \item $\epsilon(t_{\rm E})$ is the detection efficiency from \cite{2017Natur548183M}
    \item $f_A$ represents the area fraction covered by observations
\end{itemize}
The detection efficiency $\epsilon(t_{\rm E})$, taken from the extended data tables of \cite{2017Natur548183M}, is shown in Figure~\ref{fig:detection_efficiency}. Recent research shows that false positives could have an impact on the detection efficiency \cite{CrispimRomao:2025kxx}, we will study this effect in our future work.
Using the detection efficiency data shown in Figure~\ref{fig:detection_efficiency} and the expected event count formula (Eq.~\ref{eq:expected_number}), we can perform a quantitative comparison with the OGLE observational data. 
\begin{figure}[t]
    \centering
    \includegraphics[width=0.9\linewidth]{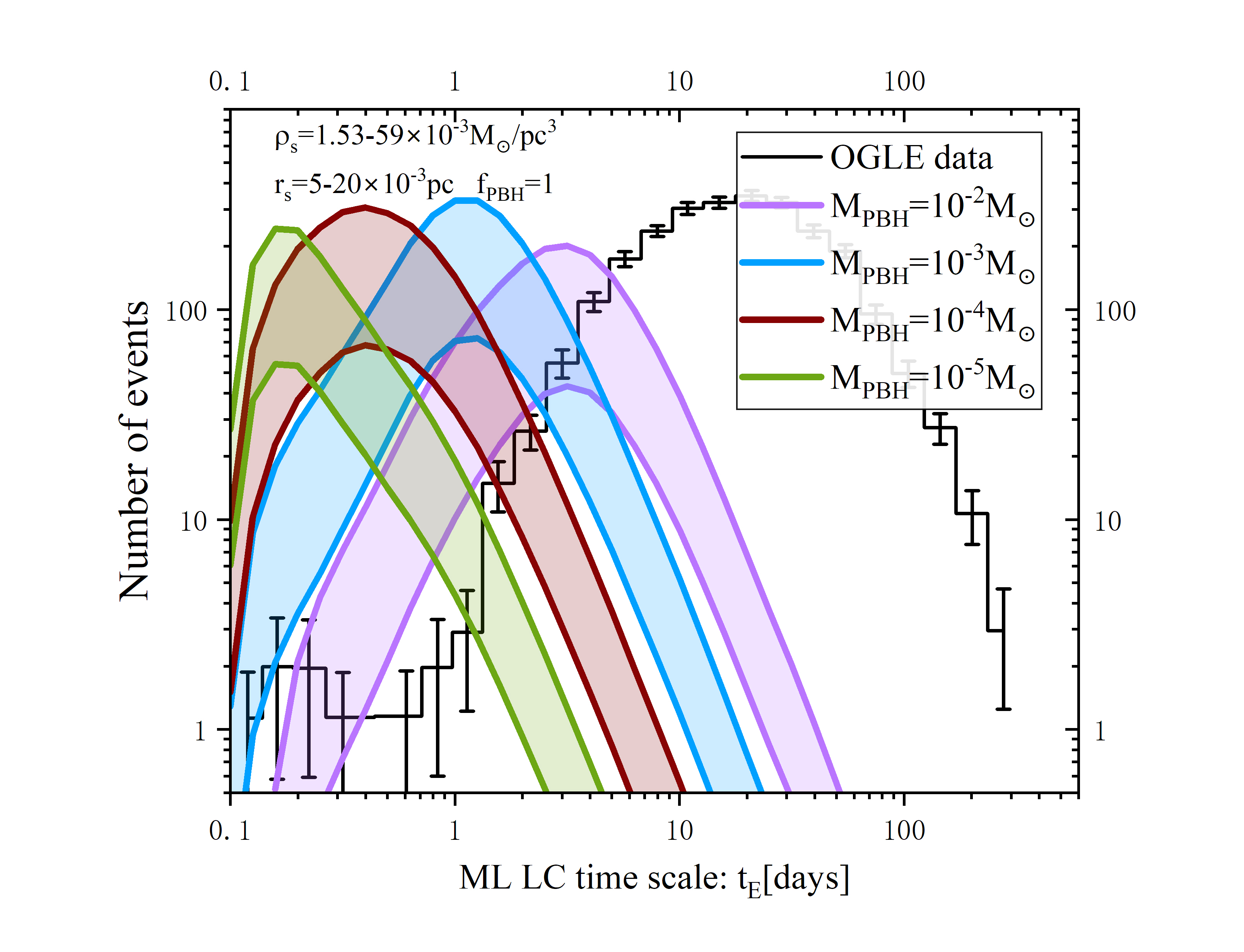}
    \caption{Comparison with five years OGLE data for Einasto profile. The istogram with errorbars denotes the OGLE data. The other colorful solid lines represent the predictions with various mass for the PBH, and we have set $f_{\rm PBH}=1$ and other parameters are the same with fig. \ref{fig:detection_efficiency}. }
    \label{fig:comparison with ogle data for Einasto profile}
\end{figure}

\subsubsection{Numeric of Einasto profile}
\label{numeric of Einasto profile}
Fig.~\ref{fig:comparison with ogle data for Einasto profile} compares the Einasto profile predictions with five-year OGLE data under the assumption $f_{\rm PBH} = 1$. Crucially, we note that:
\begin{itemize}
    \item For $M_{\rm PBH} > 10^{-1}M_\odot$, the peak amplitude decreases monotonically with increasing PBH mass
    \item For $M_{\rm PBH} < 10^{-6}M_\odot$, the peak position shifts progressively leftward
\end{itemize}
The analysis explicitly demonstrates that the predicted event rate \textit{cannot} reconcile with OGLE observations across the full PBH mass range. \textbf{We therefore conclude that models wherein dark matter consists entirely of Einasto-profile PBHs are ruled out.}

However, the Einasto profile remains viable when $f_{\rm PBH} < 1$. Our strategy involves fitting the observed peak in OGLE data at $t_{\rm E} = 0.3\,\mathrm{days}$ using this profile. 


\begin{figure}
\begin{minipage}{0.52\linewidth}
\vspace{0.2pt}
\centerline{\includegraphics[width=\textwidth]{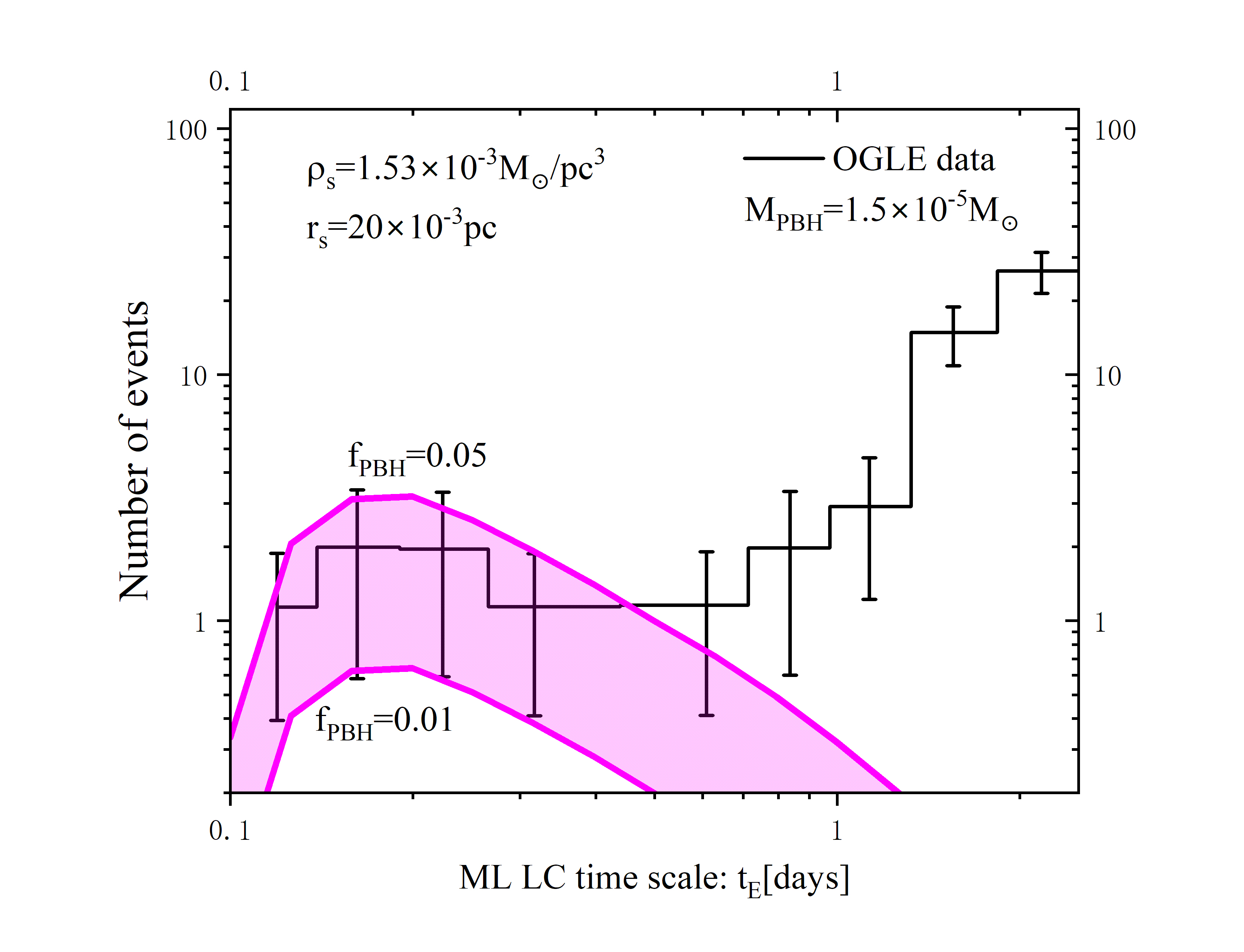}}\vspace{2pt}
\end{minipage}
\hspace*{\fill}
\begin{minipage}{0.52\linewidth}
\vspace{0.2pt}
\centerline{\includegraphics[width=\textwidth]{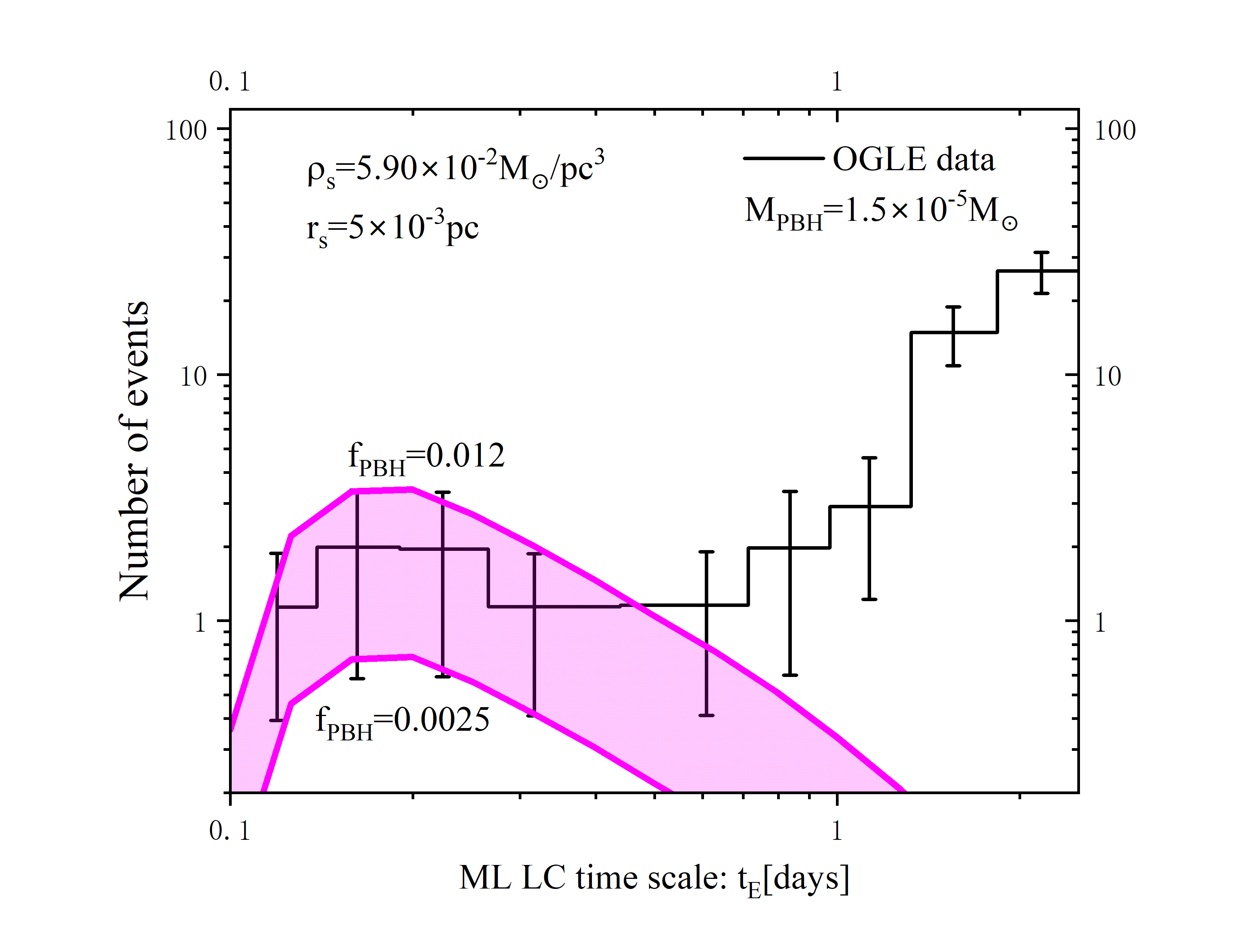}}\vspace{2pt}
\end{minipage}
\caption{Best-fit to OGLE data for the Einasto profile. 
The parameter with different central densities can range from $f_{\rm PBH} = 0.0025$ to $f_{\rm PBH} = 0.05$ represent the lower and upper bounds for $f_{\rm PBH}$. The timescale spans $0.1$ to $1.8$ days. Other parameters remain consistent with previous figures. The black histogram represents OGLE data of microlensing event number per logarithmic bin of $t_E$}
\label{fig:best fit with ogle data for Einasto profile}
\end{figure}

Fig.~\ref{fig:best fit with ogle data for Einasto profile} demonstrates the optimal fit to OGLE data at $t_{\rm E} = 0.3\,\mathrm{days}$, although depends on the scale density, but can be only up to around 5 percentages.
The error bars in the histogram represent statistical deviations per $t_{\rm E}$ bin. From this analysis, we constrain the PBH fraction to $0.0025 < f_{\rm PBH} < 0.05$, indicating that Einasto-profile PBHs can constitute only a minor component of dark matter.

\subsubsection{Numeric for Burkert Profile}
\label{numeric of burkert profile}
\begin{figure}
\begin{minipage}{0.52\linewidth}
\vspace{0.2pt}
\centerline{\includegraphics[width=\textwidth]{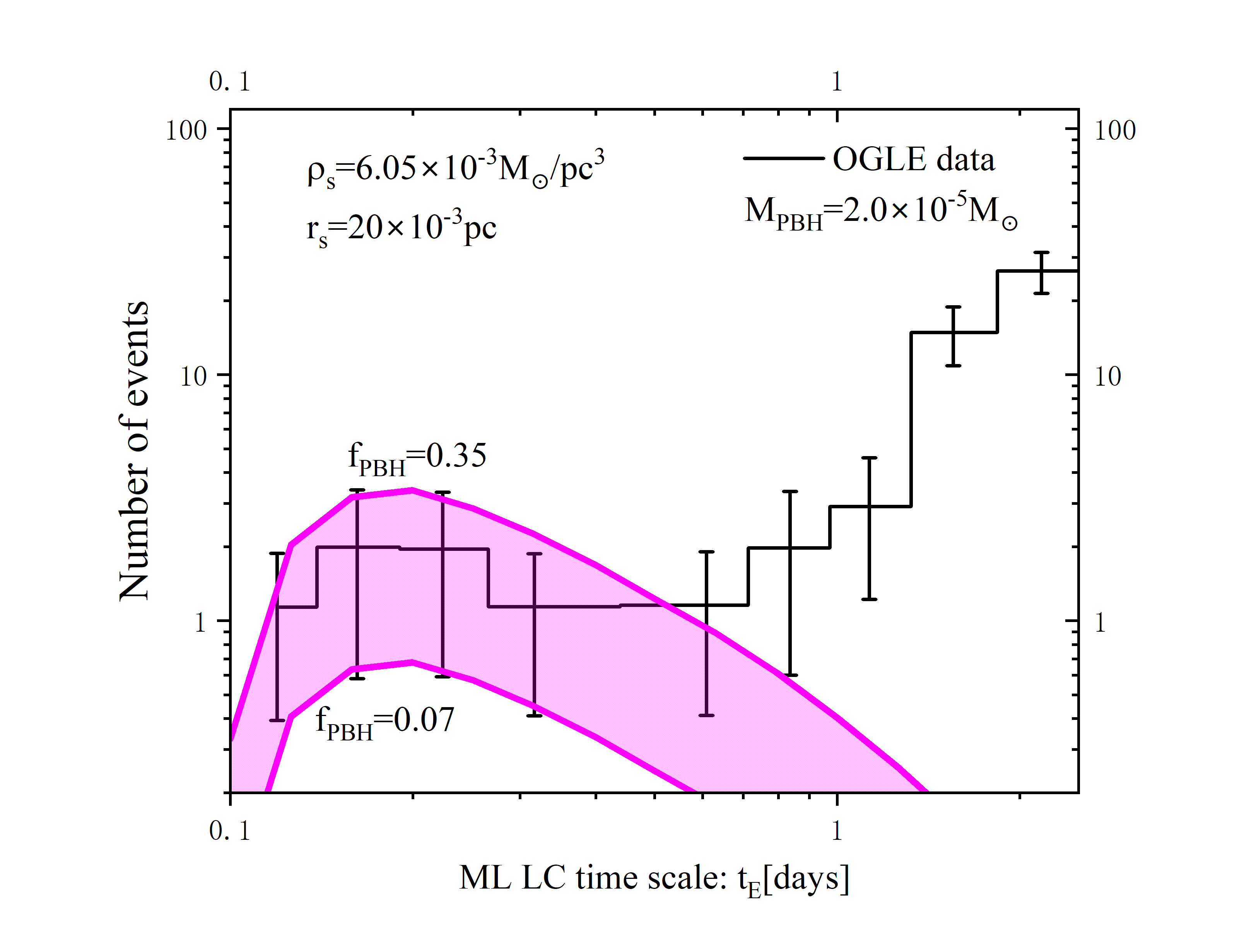}}\vspace{2pt}
\end{minipage}
\hspace*{\fill}
\begin{minipage}{0.52\linewidth}
\vspace{0.2pt}
\centerline{\includegraphics[width=\textwidth]{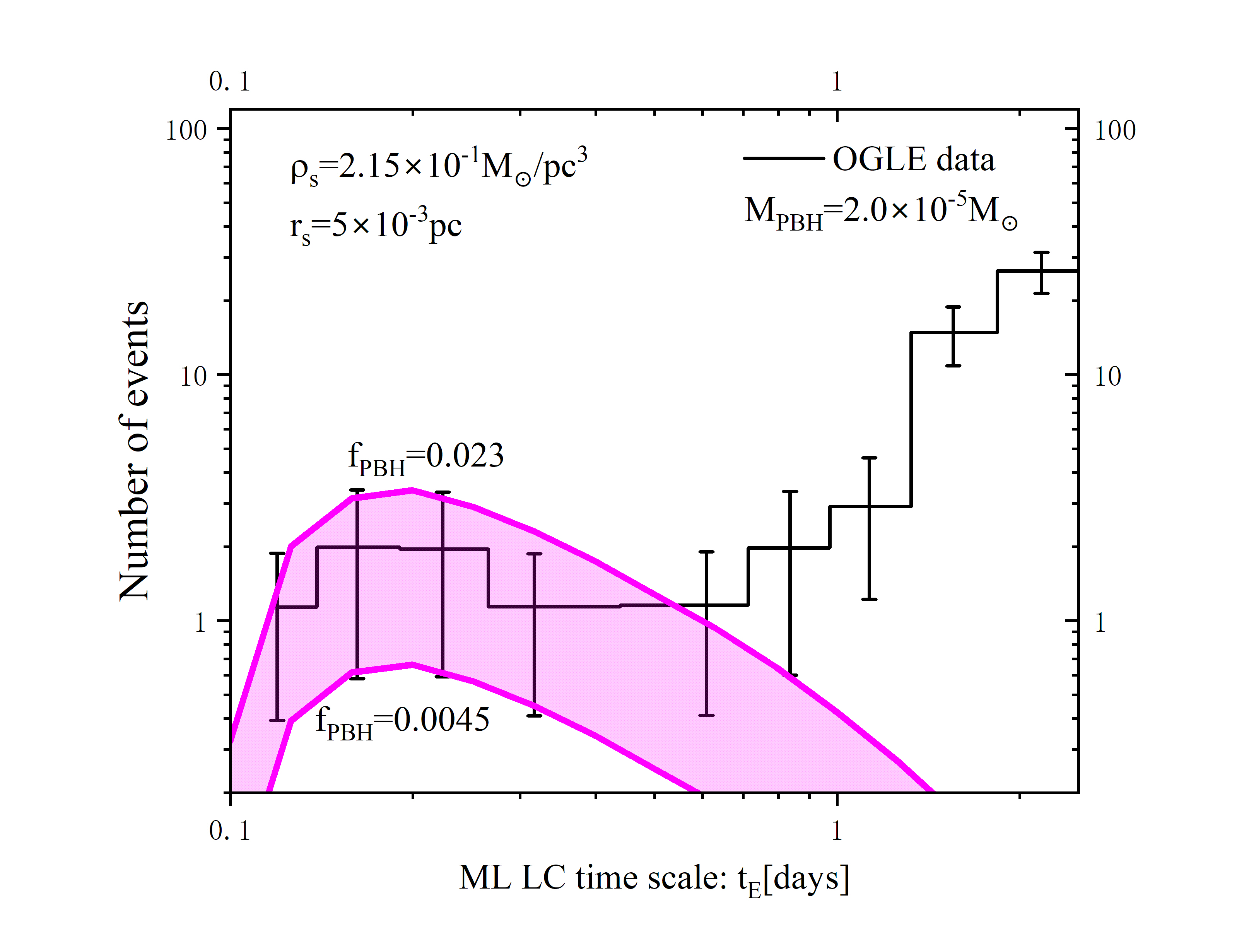}}\vspace{2pt}
\end{minipage}
\caption{Best-fit to OGLE data for the Burkert profile. The expected number of microlensing event within a time scale $t_E-\Delta t_E/2<t<t_E+\Delta t_E/2$ ($0.1$ to $1.8$ days). The five subfigures correspond to five different structural parameters in Fig.~\ref{fig:event_rate_for_Burkert_profile}. The plot shows the the range of $f_{\rm PBH}$ is from $0.0045$ to $0.35$. The black histogram represents OGLE data of microlensing event number per logarithmic bin of $t_E$.}
\label{fig:event_number_for_Burkert_profile}
\end{figure}

The comparison between the 5-year OGLE observational data and our model predictions for event counts per timescale is presented in Fig.~\ref{fig:event_number_for_Burkert_profile}, assuming a Burkert profile for the dark matter distribution. 
Key features of the analysis include:
\begin{itemize}
    \item \textbf{Stellar background:} The majority of OGLE events are attributable to stellar microlensing \cite{Niikura:2019kqi}. We therefore focus on the six ultrashort-timescale events within $t_E \in [0.1, 0.3]\,\mathrm{days}$.
    
    \item \textbf{PBH signal characteristics:} Our model predicts:
    \begin{itemize}
        \item A consistent peak at $t_E \sim 0.2\,\mathrm{days}$ across all structural parameters
        \item A steeper decline at shorter timescales, matching the OGLE data trend
    \end{itemize}
    
    \item \textbf{$f_{\rm PBH}$ scaling:} Equation~\ref{optical depth for pbh} establishes $\tau_{\rm PBH} \propto f_{\rm PBH}$. Consequently, the PBH microlensing event rate (Eq.~\ref{eq:diff_event_rate}), being proportional to $\tau_{\rm PBH}$, scales linearly with $f_{\rm PBH}$ (demonstrated in Fig.~\ref{fig:event_number_for_Burkert_profile}).
    
    \item \textbf{Parameter dependence:} The Burkert profile permits higher PBH fractions while maintaining consistency with observations. For example, with $\rho_s = 6.05 \times 10^{-3} M_\odot \mathrm{pc}^{-3}$ and $r_0 = 20 \times 10^3 \mathrm{pc}$, PBHs of mass $M_{\rm PBH} = 2.0 \times 10^{-5} M_\odot$ can constitute 35\% of dark matter.
\end{itemize}

In this section, we systematically evaluate microlensing event counts for the Einasto profile (Eq.~\eqref{profile of Einasto}) and Burkert profile (Eq.~\eqref{profile of Burkert})
with parameters consistent with kinematic data.
We demonstrated that, similar to the situation of NFW profile \cite{Niikura:2019kqi}, the microlensing events due to PBHs residing in a Einasto or Burkert profile can reproduce the ultrashort-timescale events in OGLE data. 
For the Einasto profile with monochromatic PBH $M_{\mathrm{PBH}} = 1.5 \times 10^{-5} M_{\odot}$, the PBH fraction can vary from $0.0025$ to $0.05$, depending on the scale density. 
For Burkert profile, the best-fit PBH mass is $M_{\mathrm{PBH}} = 2.0 \times 10^{-5} M_{\odot}$. Burkert profile allows a higher PBH fraction, ranging from $0.0045$ to $0.35$.
However, these constraints apply only to fixed PBH masses in select cases. 
To map the full parameter space where $f_{\mathrm{PBH}}$ varies with $M_{\mathrm{PBH}}$, a statistical treatment is required. In the following section, we implement Poissonian likelihood analysis to rigorously constrain $f_{\mathrm{PBH}}$.

\section{Constraints on PBH abundance from OGLE microlensing Data}
\label{constraints of PBH abunance}
In this section, we further constrain the PBH abundance, denoted by $f_{\mathrm{PBH}}$, using five years of microlensing data from the OGLE survey. Following the methodology of Ref.~\cite{Niikura:2017zjd}, we adopt the \textit{null hypothesis} that no microlensing events due to PBHs are observed in the OGLE dataset. This conservative assumption could give the most stringent upper limits on $f_{\mathrm{PBH}}$.

Assuming that microlensing events at different timescales are statistically independent, the number of events observed in each timescale bin follows a Poisson distribution. Specifically, the probability of observing $k$ microlensing events in a given bin is
\begin{equation}
P(k) = \frac{\lambda^k e^{-\lambda}}{k!} ~,
\end{equation}
where $\lambda$ denotes the expected number of events in that bin.

Let the total number of timescale bins be $n_\mathrm{bin}$ (with $n_\mathrm{bin} = 25$ based on OGLE data~\cite{Niikura:2019kqi}), and label the timescale of the $i$-th bin as $t_{E,i}$. Denote the observed number of events at $t_{E,i}$ as $N_{\mathrm{obs}}(t_{E,i})$, and the expected number of PBH-induced events (computed from Eq.~(\ref{eq:expected_number})) as $N_{\mathrm{exp}}^{\mathrm{PBH}}(t_{E,i})$. Then, under the null hypothesis, the total expected number of events in the $i$-th bin is
\begin{equation}
\lambda\left(t_{E,i}\right) = N_{\mathrm{obs}}\left(t_{E,i}\right) + N_{\mathrm{exp}}^{\mathrm{PBH}}\left(t_{E,i}\right)~.
\end{equation}
This expression assumes that each lensing object produces at most one observable event, which holds in the regime of low optical depth ($\tau \ll 1$), applicable to our case. 
Then the total log likelihood of all timescale bin $t_{E,i}$ is given by
\begin{equation}
\ln L(\mathbf{d}|\boldsymbol{\theta}) = \sum_{i=1}^{n_\mathrm{bin}} \left[ N_{\mathrm{obs}}(t_{E,i}) \ln \lambda(t_{E,i}) - \lambda(t_{E,i}) - \ln \left(N_{\mathrm{obs}}(t_{E,i})!\right) \right]~,
\end{equation}
where $\mathbf{d}$ represents the data vector and $\boldsymbol{\theta}$ is the model vector, including $f_{\mathrm{PBH}}$.

Under the null hypothesis, the likelihood is maximized at $f_{\mathrm{PBH}} = 0$. For a given PBH mass $\mathcal{M}$, the posterior probability distribution of $f_{\mathrm{PBH}}$ is derived via Bayes’ theorem:
\begin{equation}
P\left(f_{\mathrm{PBH}} \mid \mathbf{d}, \mathcal{M}\right) = \frac{\mathcal{L}(\mathbf{d} \mid f_{\mathrm{PBH}})\, \Pi(f_{\mathrm{PBH}})}{P(\mathbf{d} \mid \mathcal{M})}~,
\label{posterior_fPBH}
\end{equation}
where $\Pi(f_{\mathrm{PBH}})$ is the prior on $f_{\mathrm{PBH}}$, and $P(\mathbf{d} \mid \mathcal{M})$ is the Bayesian evidence. We assume a flat prior, $\Pi(f_{\mathrm{PBH}}) = \mathrm{const}$. The 95\% confidence level (C.L.) upper limit on $f_{\mathrm{PBH}}$ is then obtained by solving
\begin{equation}
\int_{0}^{f_{\mathrm{PBH},\,95\%}} \mathrm{d} f_{\mathrm{PBH}}\, P(f_{\mathrm{PBH}} \mid \mathbf{d}, \mathcal{M}) = 0.95~.
\end{equation}

\begin{figure}[t]
    \centering
    \includegraphics[width=1.1\linewidth]{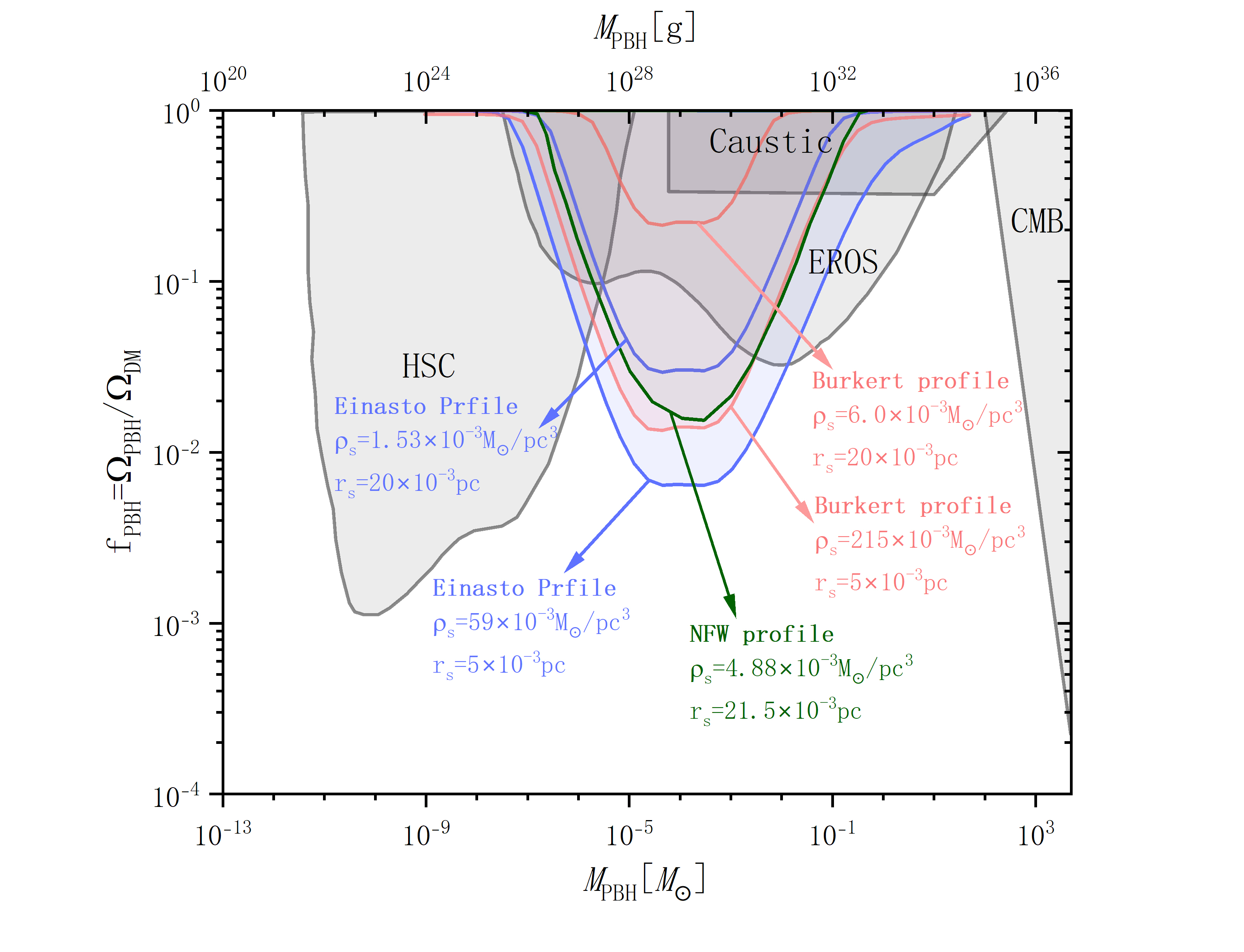}
    \caption{
    This figure shows the 95\% C.L. upper bounds on the PBH dark matter fraction $f_{\mathrm{PBH}}$ assuming a monochromatic mass spectrum and no PBH microlensing events in OGLE data. The blue and red curves correspond to constraints derived using Einasto and Burkert profiles, respectively. The green curve shows the result from Ref.~\cite{Niikura:2019kqi}, which adopted a Navarro-Frenk-White (NFW) profile. 
    Additional constraints from other observations are shown in gray: Subaru HSC~\cite{Niikura:2017zjd}, MACHO/EROS/OGLE~\cite{2007A&A...469..387T}, caustic microlensing~\cite{Oguri:2017ock}, and CMB accretion limits~\cite{Ali-Haimoud:2016mbv}.
    }
    \label{fig:constraint_fPBH}
\end{figure}

This procedure yields the upper limit on the PBH abundance $f_{\mathrm{PBH}}$ at a fixed PBH mass $\mathcal{M}$, based on the OGLE microlensing observations. The results computed with Einasto profile and Burkert profile are shown in Fig.~\ref{fig:constraint_fPBH}.

In Fig.~\ref{fig:constraint_fPBH}, we present the 95\% confidence level (C.L.) upper bounds on the PBH abundance, $f_{\rm PBH}$, across a mass range of $M_{\rm PBH} \sim [10^{-11}, 10^3]\,M_\odot$. The blue and red shaded regions represent the excluded parameter space assuming Einasto and Burkert DM halo profiles, respectively, while the green curve corresponds to the constraint assuming the NFW profile \cite{Niikura:2019kqi}.
As shown in Fig.~\ref{fig:constraint_fPBH}, a higher scale density \(\rho_s\) leads to a more stringent constraint on \(f_{\mathrm{PBH}}\). For the Einasto profile, the constraint on \(f_{\mathrm{PBH}}\) at \( M_{\mathrm{PBH}} \sim 10^{-4}\, M_\odot \) ranges from 0.0065 to 0.030, depending on the parameters. For the Burkert profile, the PBH fraction is constrained to \( f_{\mathrm{PBH}} \simeq 0.014\!-\!0.22 \). The constraint derived from the NFW profile lies between the strongest and weakest limits of the Einasto and Burkert profiles and is more consistent with the Einasto case.  


The different central densities can affect the predicted microlensing event rate through Eq.~\ref{differential event rate for pbh}. For a fixed PBH mass and fraction \(f_\mathrm{PBH}\), a dark matter halo that encloses more mass within the source distance \(D_S\) leads to a greater number of expected lensing events, as more PBHs would be present along the line of sight.

\section{Summary and outlook}
\label{Summary}
In this study, we utilized five years of OGLE microlensing data to constrain the abundance of PBHs under two DM density profiles: the Einasto and Burkert profiles. The OGLE dataset contains over 2,000 microlensing events, providing a rich dataset for probing lens populations. These events arise from various compact objects, including brown dwarfs, main-sequence (MS) stars, and stellar remnants such as white dwarfs, neutron stars, and astrophysical black holes, which can explain the main peak in the OGLE data. Notably, OGLE also reports six ultrashort-timescale events with Einstein crossing times \( t_E \sim 0.1{-}0.3 \) days. Ref.~\cite{Niikura:2019kqi} shows that, under the assumption of monochromatic mass spectrum and DM distribution with NFW profile, these events could be interpreted as PBH microlensing signals. Following Ref.~\cite{Niikura:2019kqi}, we explore how replacing the NFW profile with the Einasto or Burkert profiles affects the inferred constraints on the PBH abundance. 
The parameters are chosen by assuming virial mass $M_\mathrm{vir}=10^{12} \mathrm{M}_\odot$ and consisting with the kinematic data \cite{2025NewAR.10001721H}. Then we numerically compute the differential event rates under both profiles and compare them with the expected rate from MS stars.

For the Einasto profile, Fig.~\ref{fig:differential event rate fo Einasto} shows the differential event rate across a range of PBH masses. None of the PBH models reproduce the shape or amplitude of the MS star distribution (with \( 0.8 M_\odot < M < 1 M_\odot \)). 
A direct comparison with OGLE data (Fig.~\ref{fig:comparison with ogle data for Einasto profile}) indicates that a fully PBH-dominated DM model (\( f_{\mathrm{PBH}} = 1 \)) is inconsistent with observations. However, six ultrashort-timescale events in OGLE can be explained by PBHs with \( M_{\mathrm{PBH}} = 1.5 \times 10^{-5} M_\odot \) and \( f_{\mathrm{PBH}} = 0.019 \), suggesting PBHs could constitute a small fraction of DM (Fig.~\ref{fig:best fit with ogle data for Einasto profile}). 
As with the Einasto case, differential event rate of the Burkert profile fails to match the MS star signal. 
When compared with the OGLE data (Fig.~\ref{fig:event_number_for_Burkert_profile}), particularly the subset of ultrashort-timescale events, the inferred PBH abundance 
is higher than that from Einasto profile, indicating that the DM halo with Burkert profile can have more PBHs as its component.
Taken together, Figs.~\ref{fig:event_number_for_Burkert_profile} and~\ref{fig:best fit with ogle data for Einasto profile} imply that PBHs may offer a viable explanation for the observed population of free-floating planets.

To place upper limits on \( f_{\mathrm{PBH}} \) as a function of \( M_{\mathrm{PBH}} \), we apply a Poisson likelihood analysis under the null hypothesis that no PBH microlensing events are observed in the OGLE dataset. The resulting 95\% confidence level upper bounds are shown in Fig.~\ref{fig:constraint_fPBH} as shaded exclusion regions. 
The constraints derived from the NFW profile \cite{Niikura:2019kqi} lie between the strongest and weakest limits obtained from the Einasto and Burkert profiles. 
For a given profile, the constraint depends on its scale density. In the case of the Einasto profile, PBHs with monochromatic masses in the range \( M_{\mathrm{PBH}} \simeq 10^{-5} \!-\! 10^{-7}\, M_\odot \) can constitute at most 3\% of the total dark matter. 
In contrast, the Burkert profile allows a higher PBH fraction. With certain parameters, PBHs can contribute up to 22\% of the total dark matter.


Our findings demonstrate that constraints on the PBH abundance are sensitive to the assumed DM density profile, emphasizing the importance of accurate modeling of the Galactic halo—particularly the inner region. Thus, improving our understanding of the central DM distribution is important for robust PBH constraints. This study focuses solely on gravitational lensing effects and does not consider additional physics such as PBH clustering~\cite{Clesse:2016vqa}, dynamical heating, or interactions with other DM candidates like axions or WIMPs~\cite{Fairbairn:2017sil, Fairbairn:2017dmf, Carr:2020mqm}. Furthermore, dark matter halo formation is complex \cite{Carr:2020mqm}; the resulting profile depends on factors including:
\begin{itemize}
    \item The initial PBH fraction in dark matter
    \item The spatial distribution of PBHs
    \item Galactic tidal forces
\end{itemize}
Additionally, PBHs may exist in isolated or clustered configurations, which would also affect microlensing event rates. Our methodology can be extended to other dark matter candidates such as axions or WIMPs. Future work could also explore:
\begin{itemize}
    \item Finite-source effects in microlensing \cite{Cai:2022kbp}
    \item Lens size dependencies \cite{Fujikura:2021omw}
    \item Other microlensing parameters \cite{Carr:2025kdk}
\end{itemize}

\acknowledgments
LH, BC and CY are supported by the National Natural Science Foundation of China (Grant No.~12165009) and the Hunan Natural Science Foundation (Grants No.~2023JJ30487 and No.~2022JJ40340). LH gratefully acknowledges early-stage discussions with Tomislav Prokopec on this project and thanks him for his guidance in cosmology. ZR is supported from the National Natural Science Foundation of China under Grant 12305104 and the Education Department of Hunan Province under Grant No. 24B0503.

\bibliography{microlensing_milky_way}
\bibliographystyle{JHEPmod}
\end{document}